\documentclass[a4paper,11pt]{article}

\usepackage[english]{babel}
\usepackage[utf8]{inputenc}
\usepackage[T1]{fontenc}

\usepackage{amsmath,amsfonts,amsthm,amssymb}
\usepackage{manfnt}
\usepackage{stmaryrd}
\usepackage{subcaption}
\usepackage{graphicx}
\usepackage{proof}
\usepackage{tikz}
\usepackage{mathtools}

\usepackage{hyperref}
\usepackage{cite}

\usepackage{lineno}

\graphicspath{{./img/}}

\renewcommand{\phi}{\varphi}
\renewcommand{\epsilon}{\varepsilon}

\newcommand{\tileset}{\ensuremath{\mathbf{T}} }
\newcommand{\prototile}{\mathbf{t}}

\newcommand{\tiling}{\ensuremath{\mathcal{T}} }
\newcommand{\lang}[1]{\ensuremath{\mathcal{L}(#1)} }

\newcommand{\fullshift}{X_\tileset}
\newcommand{\shift}{\mathfrak{S}}
\newcommand{\ie}{\emph{i.e.} }

\newcommand{\tendsto}{\underset{n\to\infty}{\longrightarrow}}
\newcommand{\dsa}{\mathbf{d}_1} 
\newcommand{\dsb}{\mathbf{d}_2} 
\newcommand{\dwc}{\mathrm{d}_3} 
\newcommand{\dwd}{\mathrm{d}_4} 

\newcommand{\nN}{{n \in \mathbb{N}} }

\newcommand{\ball}[1]{\mathcal{B}_{#1}}
\newcommand{\sphere}[1]{\mathcal{S}_{#1}}
\newcommand{\radiuses}[2]{\mathrm{r}_{#1}(#2)}
\newcommand{\boundary}{\partial}

\newcommand{\support}[1]{\mathrm{supp}(#1)}

\renewcommand{\vec}[1]{\mathbf{#1}} 

\newtheorem{theorem}{Theorem}
\newtheorem{definition}{Definition}
\newtheorem{proposition}{Proposition}
\newtheorem{lemma}{Lemma}

\newtheorem{remark}{Remark}
\newtheorem{example}{Example}

\title{Geometrical tilings : distance, topology, compactness and completeness}
\author{Victor H. Lutfalla \footnote{Université Publique, Université Aix-Marseille, Université de Caen}}
\date{\today}
\begin{document}

\maketitle

\paragraph{Abstract.}
We present the different distances on tilings of $\mathbb{R}^d$ that exist in the literature, we prove that (most of) these definitions are correct (\ie they indeed define metrics on tilings of $\mathbb{R}^d$). We prove that for subshifts with finite local complexity (FLC) these metrics are topologically equivalent and even metrically equivalent, and also we present classical results of compactness and completeness.
Note that, excluding the equivalence of these metrics, all of the results presented here are known (see for example the survey \cite{robinson2004}) however we were unable to find a reference with complete proofs for some of these results so we decided to write this notice to clarify some definitions and give full proofs.



\section{Definitions and main results}
\label{sec:def}
Let $d$ be an positive integer. Usually we consider $d\in\{2,3\}$ but the definitions and results apply for any positive integer.

A \emph{tile} is a compact of $\mathbb{R}^d$ which is the closure of its interior. A \emph{prototile} is a tile up to translation. We denote by $\equiv$ the up-to-translation equivalence relation.
A \emph{tiling} is a covering of $\mathbb{R}^d$ by tiles that do not overlap.

A \emph{tileset} is a finite set of prototiles.
Given a tileset $\tileset$, the set of all tilings where the tiles are translates of the prototiles of $\tileset$ is called the \emph{full-shift} and is denoted by $\fullshift$. Note that in this paper we always assume that the full-shift $\fullshift$ is not empty, note also that the Domino problem which takes as input a tileset $\tileset$ and outputs \texttt{true} if $\fullshift$ is non-empty and \texttt{false} otherwise is undecidable \cite{berger1966}.

A \emph{patch}, usually denoted by $P$, of a tiling $\tiling$ is a finite and simply connected set of tiles in the tiling. We write $P\subset \tiling$. A \emph{pattern}, denoted by $\mathcal{P}$, is a patch up to translation.
The \emph{support of a patch} is the union of its tiles which we write $\support{P}$.
The set of patterns that appear in a a tiling $\tiling$ is called \emph{language} and is denoted by $\lang{\tiling}$.

Translations act naturally on tiles and tilings. We write $t+\vec{x}$ the translate of tile $t$ by vector $\vec{x}$, and $\tiling + \vec{x}$ the translate of tiling $\tiling$ by vector $\vec{x}$ \ie $\tiling + \vec{x} := \{ t + \vec{x}, t \in \tiling\}$.

There exist various definition of distances on the tiling spaces in the literature, the general idea is always the same: two tilings are close if after a small translation they agree on a large ball. When this "agree" means that the tilings are identical on a large ball we call the distance a \emph{strong} tiling distance, when it means that they are close in terms of Hausdorff distance we call it a \emph{weak} tiling distance.

\begin{definition}[Distances on tilings]
Let \tileset be a finite set of prototiles of $\mathbb{R}^d$. Let $\fullshift$ be the set of tilings of $\mathbb{R}^d$ with tiles in \tileset. We define four distances on $\fullshift$:

\begin{align*}
&\dsa(\tiling_0,\tiling_1) := \inf \{1\} \cup \{r>0|\, \exists \vec{x}\in \ball{r}, P_0 \in \tiling_0[[\ball{1/r}]], P_1 \in \tiling_1[[\ball{1/r}]],\\
&\hspace{10cm}P_0 = P_1 +\vec{x}\}\\
&\dsb(\tiling_0,\tiling_1) := \inf \{1\} \cup \{r>0|\, \exists \vec{x}_0,\vec{x}_1\in \ball{r},
(\tiling_0+\vec{x}_0)[\ball{1/r}] = (\tiling_1+\vec{x}_1)[\ball{1/r}]\}\\
&\dwc(\tiling_0,\tiling_1) := \inf \{1\} \cup \{r>0|\,\exists P_0 \in \tiling_0[[\ball{1/r}]], P_1 \in \tiling_1[[\ball{1/r}]], H(P_0,P_1) \leq r \}\\
&\dwd(\tiling_0,\tiling_1) := \inf \{1\} \cup \{r>0|\, H( \boundary_{1/r}\tiling_0, \boundary_{1/r}\tiling_1 ) \leq r \}
\end{align*}
where $\ball{r}$ is the closed ball of centre $0$ and radius $r$ in $\mathbb{R}^d$, $\tiling[[K]]$ with $K$ a compact and $\tiling$ a tiling is the set of finite patches $P\subset \tiling$ such that $K\subset \support{P}$,
 $\tiling[K]$ with $K$ a compact and $\tiling$ a tiling is the smallest finite patch $P\in \tiling[[K]]$,
$H$ is the Hausdorff distance (see Section \ref{sec:weak} for more details),
and $\boundary_{1/r} \tiling$ is the sphere of radius $1/r$ together with the union of the portion of the boundary of the tiles of $\tiling$ that lie in $\ball{1/r}$.
\end{definition}

\begin{remark}[References and remarks on those definitions]
These four distances are used by various authors on tilings of $\mathbb{R}^d$ or similar objects such as Delone sets:
 $\dsa$ \cite{bedaride2011, robinson2004, solomyak1997dynamics},
 $\dsb$ \cite{lee2002, sadun2006},
 $\dwc$ \cite{radin1992, lenz2002, smilansky2022},
 $\dwd$ \cite{robinson1996}.\\
 Note that, for Delone sets or more generally for closed subsets of $\mathbb{R}^d$, the topology induced by the weak distance ($\dwc$ or $\dwd$) is called Chabauty-Fell topology \cite{chabauty1950, fell1962, harpe2008}.\\
 Note that in these references there might be a few differences such as using $2^{-1/2}$ as an upper bound instead of $1$, or not using an upper bound at all (in which case it does not satisfy the triangular inequality). Note also that sometimes unprecise phrases like "agree on a ball of diameter $1/\epsilon$" are used as definitions, so it might be interpreted as either $\dsa$ or $\dsb$.\\
 Note that in \cite{radin1992} they do not define a distance but directly a topology by a countable base of open sets, however the distance $\dwc$ is very close to that definition.\\
 Note that in \cite{bedaride2011} the settings are both less general since the dimension is fixed to $d=2$ and more general since they consider both tilings of the Euclidean plane and of the Hyperbolic plane.
\end{remark}

%
%

\begin{proposition}[Validity of the definitions, folk.]
These four definitions give distances on $\fullshift$ and the translation or shift action is continuous for these four distances.
\end{proposition}
Note that this is widely known, however it is hard to find a complete proof for each of those distances so we will detail the proofs in Sections \ref{sec:strong} and \ref{sec:weak}. In most references a variation on "one easily verifies that $d$ is a metric" is used, and in other references the proofs are incomplete. Note that \cite{lee2002} and \cite{bedaride2011} contain complete proofs of the triangular inequality in their respective settings.

%
%

Our main result is that these four distances are topologically and even metrically equivalent on tiling subshifts with Finite Local Complexity. The idea of Finite Local Complexity (FLC) is that a set of tilings is FLC when for any $n>0$ there are finitely many $n$-tiles patches up to translation in the set of tilings, for more details see Section \ref{sec:topological_equivalence}.

\begin{proposition}[Topological equivalence for FLC subshfits]
\label{prop:topological_equivalence}
Let \tileset be a finite set of prototiles of $\mathbb{R}^d$ such that $\fullshift$ has Finite Local Complexity (FLC).\\
The four tiling distances $\dsa$, $\dsb$, $\dwc$, and $\dwd$ are topologically equivalent \ie any convergent sequence for a distance is convergent for all the other distances.
\end{proposition}

\begin{theorem}[Metric equivalence for FLC subshifts]
\label{th:metric_equivalence}
Let \tileset be a finite set of prototiles of $\mathbb{R}^d$ such that $\fullshift$ has Finite Local Complexity (FLC).\\
The four tiling distances $\dsa$, $\dsb$, $\dwc$ and $\dwd$ are metrically equivalent \ie $\exists \alpha, \beta \in \mathbb{R}$ such that $\dsa \leq \alpha \dwc \leq \beta \dsa$ and similarly for all the pairs of distances.
\end{theorem}
Let us now recall some classical results.

\begin{theorem}[Compactness, folk.]
Let \tileset be a finite set of prototiles of $\mathbb{R}^d$.
\begin{itemize}
\item Any subshift $X\subset \fullshift$ with finite local complexity (FLC) is compact for the strong tiling metrics $\dsa$ and $\dsb$.
\item Any subshift $X\subset \fullshift$ is compact for the weak tiling metrics $\dwc$ and $\dwd$.
\end{itemize}
\end{theorem}
Note that the first item of this theorem is widely known and usually attributed to \cite{rudolph1989}, remark however that in the original paper the result is not stated as such but can be derived from the construction of section 3. For completeness we decided to present a proof of the first item using K\H{o}nig's lemma. The second item is less known, but can still be considered folk.

\begin{theorem}[Completeness, folk.]
Let \tileset be a finite set of prototiles of $\mathbb{R}^d$.
$\fullshift$ is complete for the strong tiling metrics and for the weak tiling metrics.
\end{theorem}
This result, again, is widely known, and a proof is presented in \cite{robinson2004}. For completeness we decided to write the proof nonetheless.

\paragraph{Overview of the paper.}~\\
In Section \ref{sec:strong} we define the strong tiling distances and prove that they indeed define distances on the tiling space.\\
In Section \ref{sec:weak} we define the weak tiling distances and prove that they indeed define distances on the tiling space.\\
In Section \ref{sec:topology} we take a look at the induced topology in the tiling space and we prove the continuity of the shift action.\\
In Section \ref{sec:topological_equivalence} we define finite local complexity (FLC) subshifts and present the topological equivalence of the weak and strong metrics on FLC subshifts. Note that only the main ideas of the proof are given since a stronger result is proved later.\\
In Section \ref{sec:metric_equivalence} we prove that the weak and strong metrics are metrically equivalent on FLC subshifts.\\
In Section \ref{sec:compactness} we give a proof to the classical result that FLC subshifts are compact for the strong tiling metric, we also present the less known fact that all subshifts are compact for the weak tiling metric.\\
In Section \ref{sec:completeness} we present a proof to the classical result that subshifts are complete for the tiling metrics.

%

\section{Strong distance on geometrical tilings}
\label{sec:strong}
The idea of the strong distance on tilings is that two tilings are close if after a small translation they are identical on a large ball around the origin. However the exact definition is quite tricky.

In this section we consider that a tileset $\tileset$ is fixed and that all tilings are in $\fullshift$.

We write $\ball{r}$ for the closed ball of centre 0 and radius $r$ in $\mathbb{R}^d$ \ie
\[\ball{r}:= \{\vec{x}\in\mathbb{R}^d|\ \|\vec{x}\|\leq r\}.\]


Given a tiling $\tiling$ and a compact $K$ of $\mathbb{R}^d$, we define $\tiling[[K]]$ called set of $K$-patches as the set of patches of $\tiling$ that cover $K$ \ie
\[ \tiling[[K]] := \{ P \underset{patch}{\subset}\tiling |\, K \subseteq \support{P}\}. \]
We also define the smallest $K$-patch $\tiling[K]$ as
\[\tiling[K] := \bigcap\limits_{P \in \tiling[[K]]} P.\]

\begin{lemma}[Two trivial results on $K$-patches]
\label{lemma:trivial_Kpatches}
Let $\tiling$ be a tiling, $K, K'$ be two compacts and $\vec{x}$ be a vector. We have :
\begin{itemize}
\item if $K \subseteq K'$ then $\tiling[K] \subseteq \tiling[K']$,
\item $(\tiling[K])+\vec{x} = (\tiling + \vec{x})[K + \vec{x}]$.
\end{itemize}
\end{lemma}
\begin{proof}
These results might be trivial, but for the sake of completeness let us prove them.

\emph{Inclusion.} Since $K \subseteq K'$, for any patch $P\subset \tiling$ we have $K' \subseteq \support{P} \Rightarrow K \subseteq \support{P}$ so $\tiling[[K']] \subset \tiling [[K]]$ so $\tiling[K] \subseteq \tiling[K']$. Remark however that we might have $K \subsetneq K'$ and yet $\tiling[K] = \tiling[K']$.

\emph{Translation.} we actually show that $\tiling[[K]]+\vec{x} = (\tiling + \vec{x})[[K+\vec{x}]]$. Let us take $P$ in $\tiling[[K]]$, let us show that $P+\vec{x}\in  (\tiling + \vec{x})[[K+\vec{x}]]$. Since $P$ is a patch of $\tiling$ then $P+\vec{x}$ is a patch of $\tiling+\vec{x}$, and since $P$ covers $K$ then $P+\vec{x}$ covers $K+\vec{x}$ so $P+\vec{x} \in (\tiling + \vec{x})[[K+\vec{x}]]$. Now let us take $P$ in $(\tiling + \vec{x})[[K+\vec{x}]]$, let us show that $P-\vec{x} \in \tiling[[K]]$. Since $P$ is a patch of $\tiling+\vec{x}$ then $P-\vec{x}$ is a patch of $\tiling+\vec{x}-\vec{x} = \tiling$, and since $P$ covers $K+\vec{x}$ then $P-\vec{x}$ covers $K+\vec{x}-\vec{x} = K$.
\end{proof}


From the $K$-patches we can define the radiuses of translation equivalence \ie the radiuses $r$ such that after a small translation (at most $r$) the tilings agree on a large patch (covers at least $\ball{1/r}$).

\begin{definition}[Radiuses of translation equivalence]
For $\tiling_0, \tiling_1 \in \fullshift$, we define the set of radiuses of translation equivalence of $\tiling_0$ and $\tiling_1$ as
\begin{align*}
\radiuses{1}{\tiling_0,\tiling_1} := \{ r > 0| \ &\exists \vec{x}\in\ball{r},\, \exists P_0 \in \tiling_0[[\ball{1/r}]], P_1 \in \tiling_1[[\ball{1/r}]],\\
&\hspace{6cm} P_0 = P_1 + x\}\\
\radiuses{2}{\tiling_0,\tiling_1} := \{ r > 0| \ &\exists \vec{x}_0, \vec{x}_1 \in\ball{r},\,
(\tiling_0 + \vec{x}_0)[\ball{1/r}] = (\tiling_1 + \vec{x}_1)[\ball{1/r}] \}.
\end{align*}
\end{definition}

We want to define the distance $\dsa$ and $\dsb$ as the infimum of the radiuses of translation equivalence, however this would not define distances because if two tilings are incomparable (for example disjoint subsets of tiles) the set of radiuses of translation equivalence is empty, so its infimum is $+\infty$ and therefore two elements are at infinite distance which is incompatible with the definition of a distance. Additionally, for large values the radiuses of translation equivalence do not satisfy the triangular inequality, see Examples \ref{example:broken_triangular_inequality_b} and \ref{example:broken_triangular_inequality_a}. To solve these problem we bound the distance with a carefully chosen constant. The exact value of the constant is important to have the triangular inequality.

\begin{example}[Example of tilings such that the radiuses of translation equivalence $\mathrm{r}_2$ do not satisfy the triangular inequality]
\label{example:broken_triangular_inequality_b}
For simplicity, we consider coloured tiles on $\mathbb{R}^2$. This example can be easily adapted to purely geometrical tiles.\\
$\tileset :=\{ white,\, black\}$ where $white$ and $black$ are unit square tiles of respectively white and black colour.\\
\begin{figure}[h]
\center
\begin{subfigure}{0.3\textwidth}
\includegraphics[width=\textwidth]{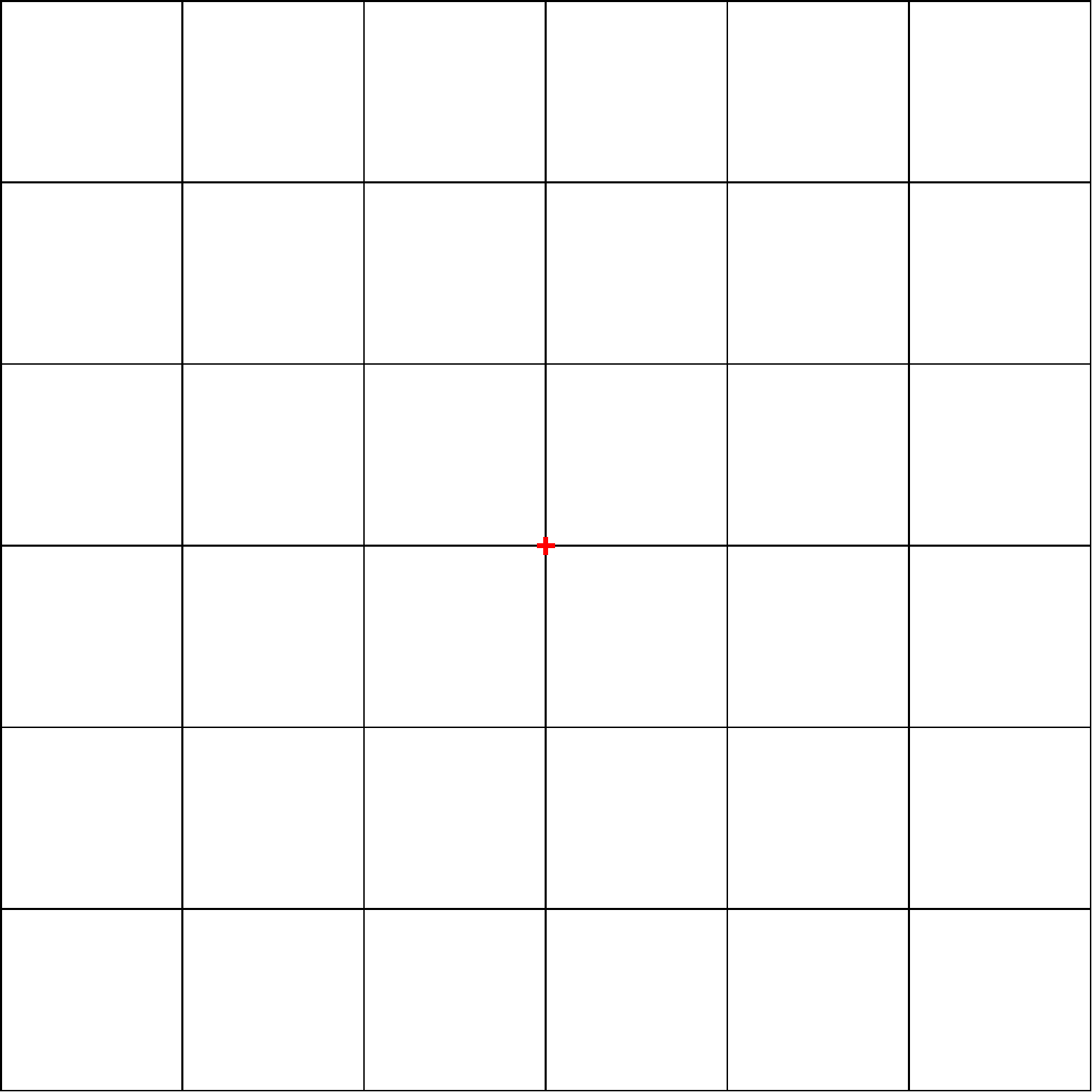}
\caption{$\tiling_{white}$}
\end{subfigure}
\hfill
\begin{subfigure}{0.3\textwidth}
\includegraphics[width=\textwidth]{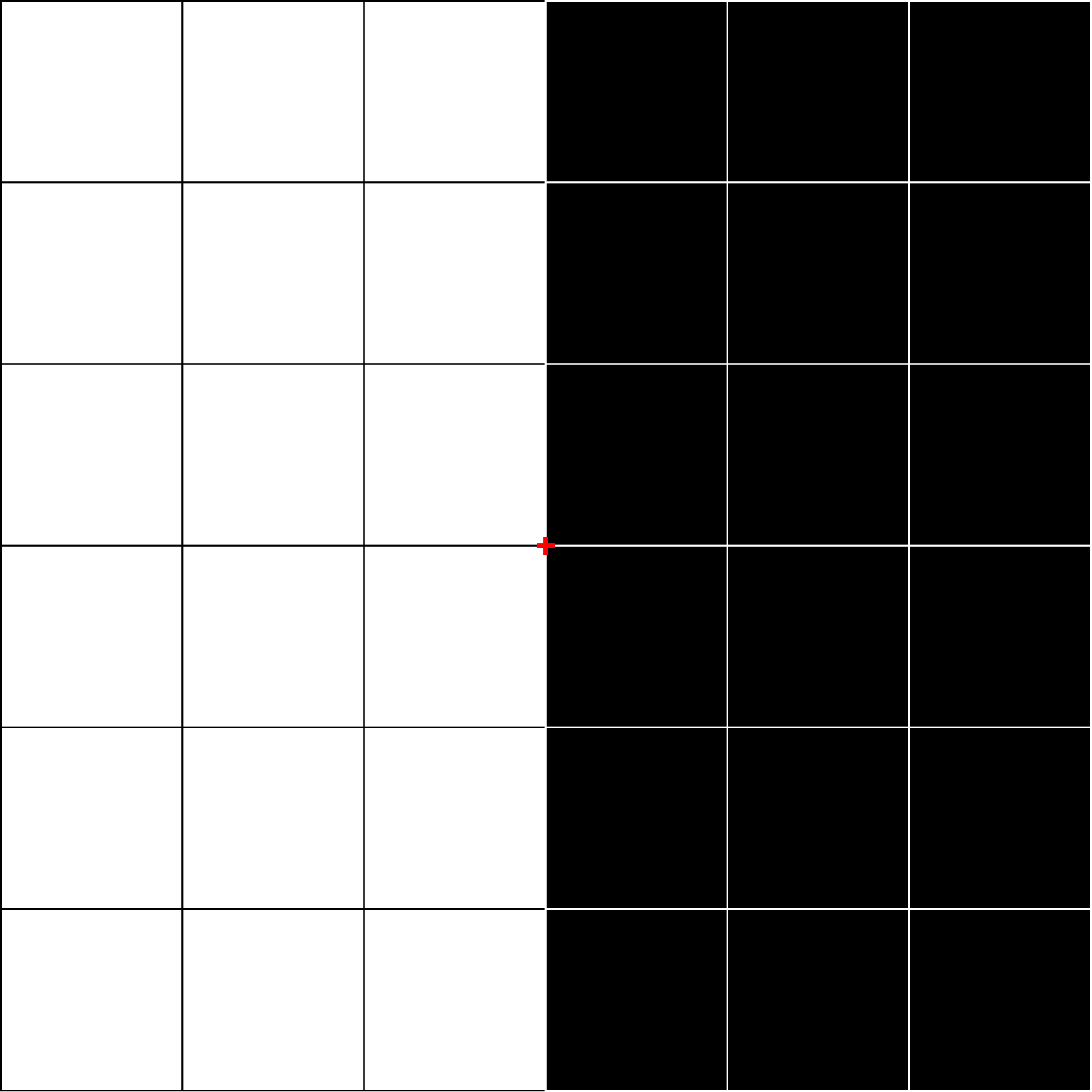}
\caption{$\tiling_{mixed}$}
\end{subfigure}
\hfill
\begin{subfigure}{0.3\textwidth}
\includegraphics[width=\textwidth]{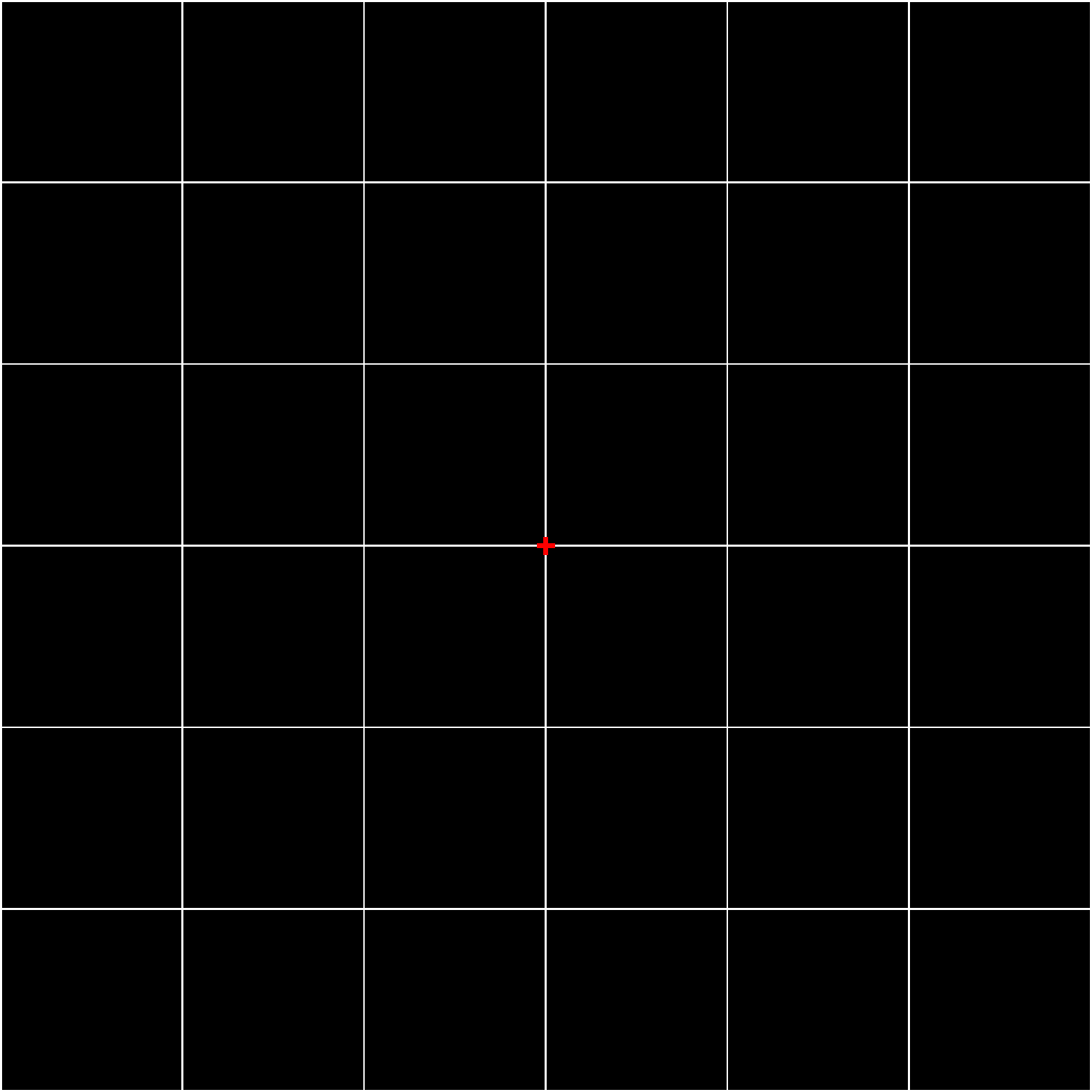}
\caption{$\tiling_{black}$}
\end{subfigure}
\caption{The central fragments of $\tiling_{white}$, $\tiling_{mixed}$ and $\tiling_{black}$.}
\label{fig:broken_triangular_inequality_b}
\end{figure}
Let $\tiling_{white}$ be the regular square tiling of white tile, $\tiling_{black}$ the regular square tiling of black tiles and $\tiling_{mixed}$ be the regular square tiling with white tiles on the left half-plane and black tiles on the right half plane, see Figure \ref{fig:broken_triangular_inequality_b}.\\
We have $\inf \radiuses{2}{\tiling_{white},\tiling_{mixed}}=1$ because $\tiling_{white}[\ball{1}] = (\tiling_{mixed} + (-1,0))[\ball{1}]$. Similarly $\inf \radiuses{2}{\tiling_{mixed},\tiling_{black}} = 1$.
However, because they have disjoint subsets of prototiles, $\tiling_{white}$ and $\tiling_{black}$ are at infinite distance \[\inf \radiuses{2}{\tiling_{white},\tiling_{black}} = +\infty > \inf \radiuses{2}{\tiling_{white},\tiling_{mixed}} + \inf \radiuses{2}{\tiling_{mixed},\tiling_{black}}.\]

Note also that by replacing a $white$ tile by a $black$ tile in $\tiling{white}$ suitably far away from the origin we can have $\inf\radiuses{2}{\tiling_{white},\tiling_{black}}$ finite and arbitrarily large.
\end{example}

\begin{example}[Example of tilings such that the radiuses of translation equivalence $\mathrm{r}_2$ do not satisfy the triangular inequality]
\label{example:broken_triangular_inequality_a}
This second example is a tiny bit more complex, for simplicity we consider coloured tiles but this can be adapted with purely geometrical tiles. The tiles are unit square with a colour.
\[\tileset:= \{ white,\, red,\, yellow \}\]

\begin{figure}[h]
\center
\begin{subfigure}{0.3\textwidth}
\includegraphics[width=\textwidth]{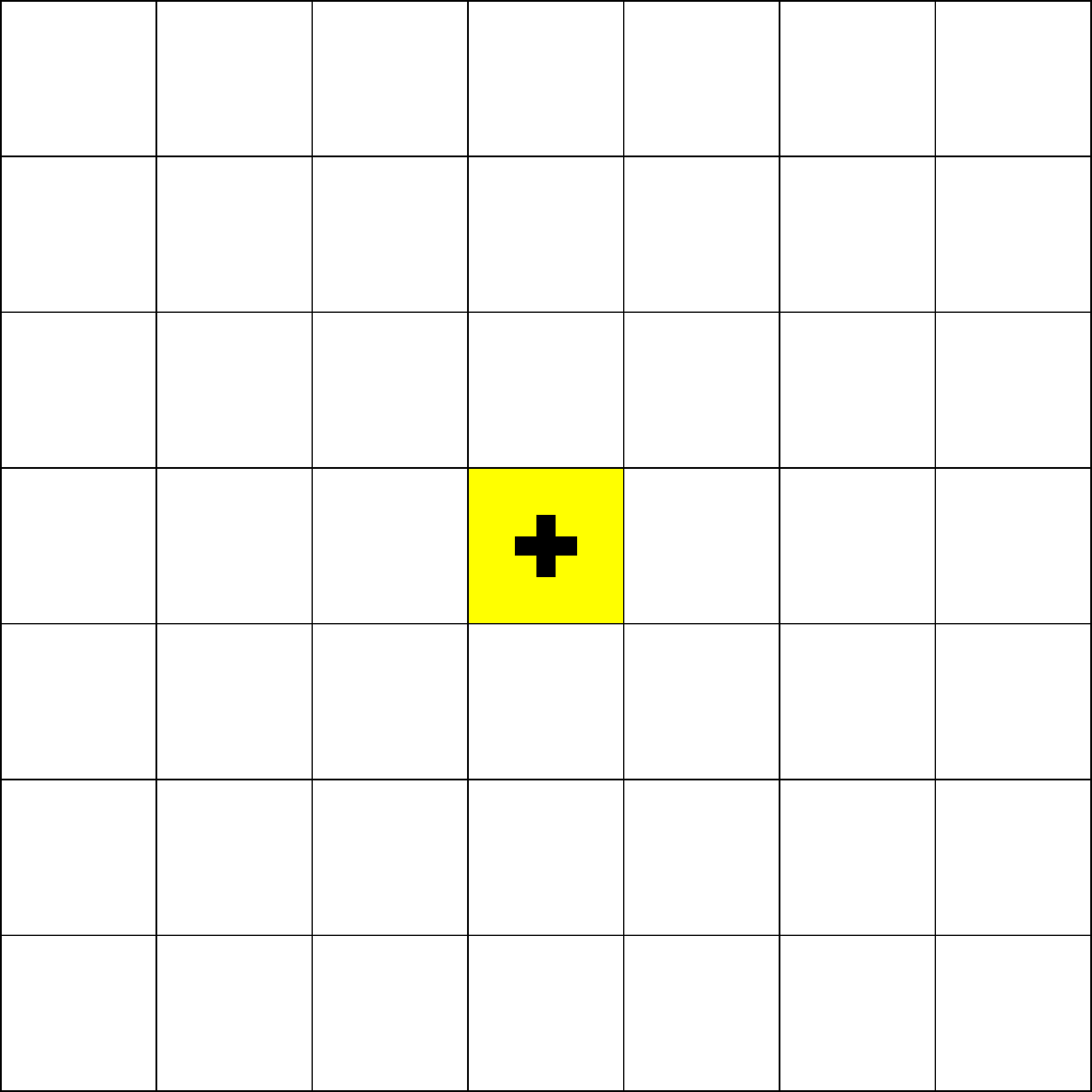}
\caption{$\tiling_{yellow}$}
\end{subfigure}
\hfill
\begin{subfigure}{0.3\textwidth}
\includegraphics[width=\textwidth]{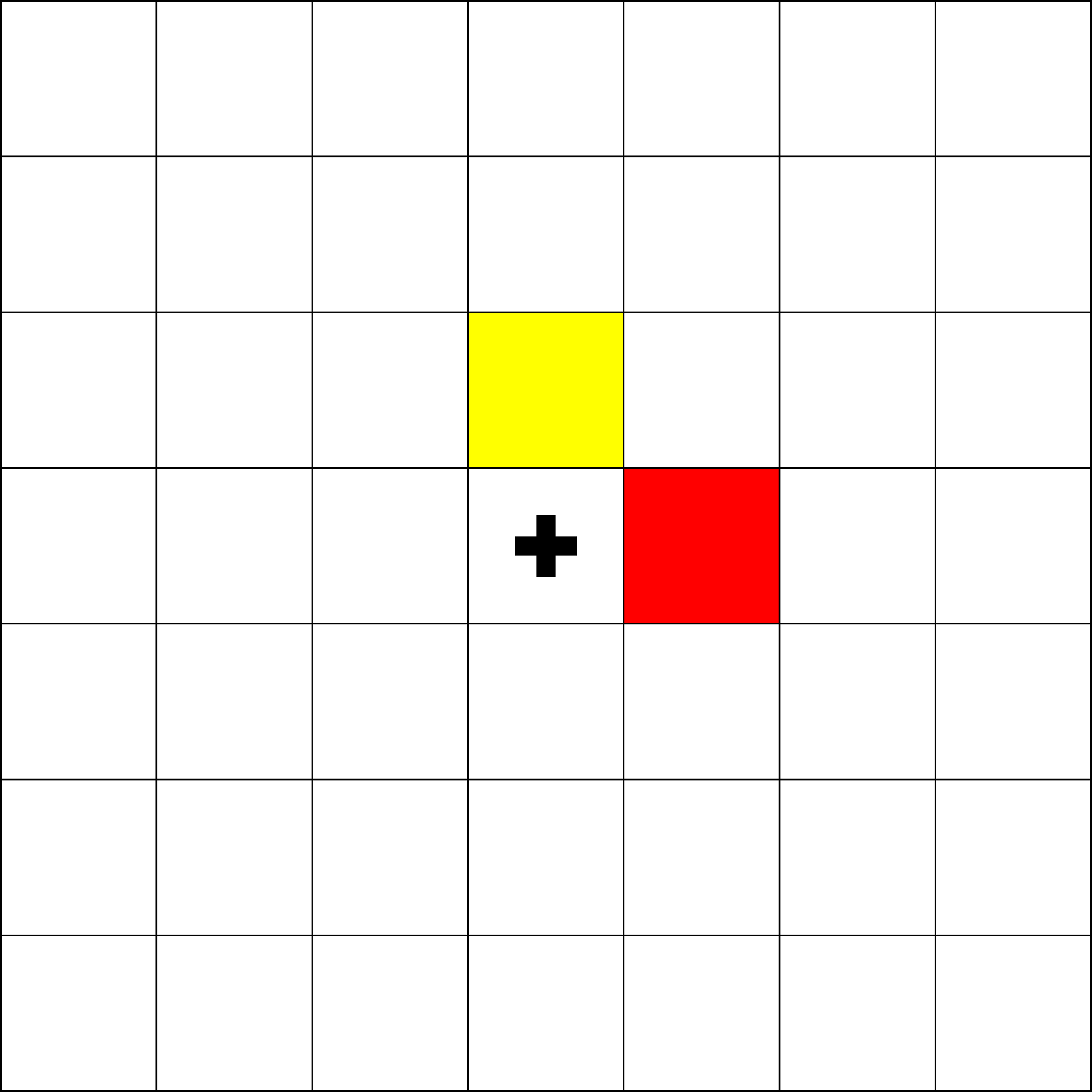}
\caption{$\tiling_{med}$}
\end{subfigure}
\hfill
\begin{subfigure}{0.3\textwidth}
\includegraphics[width=\textwidth]{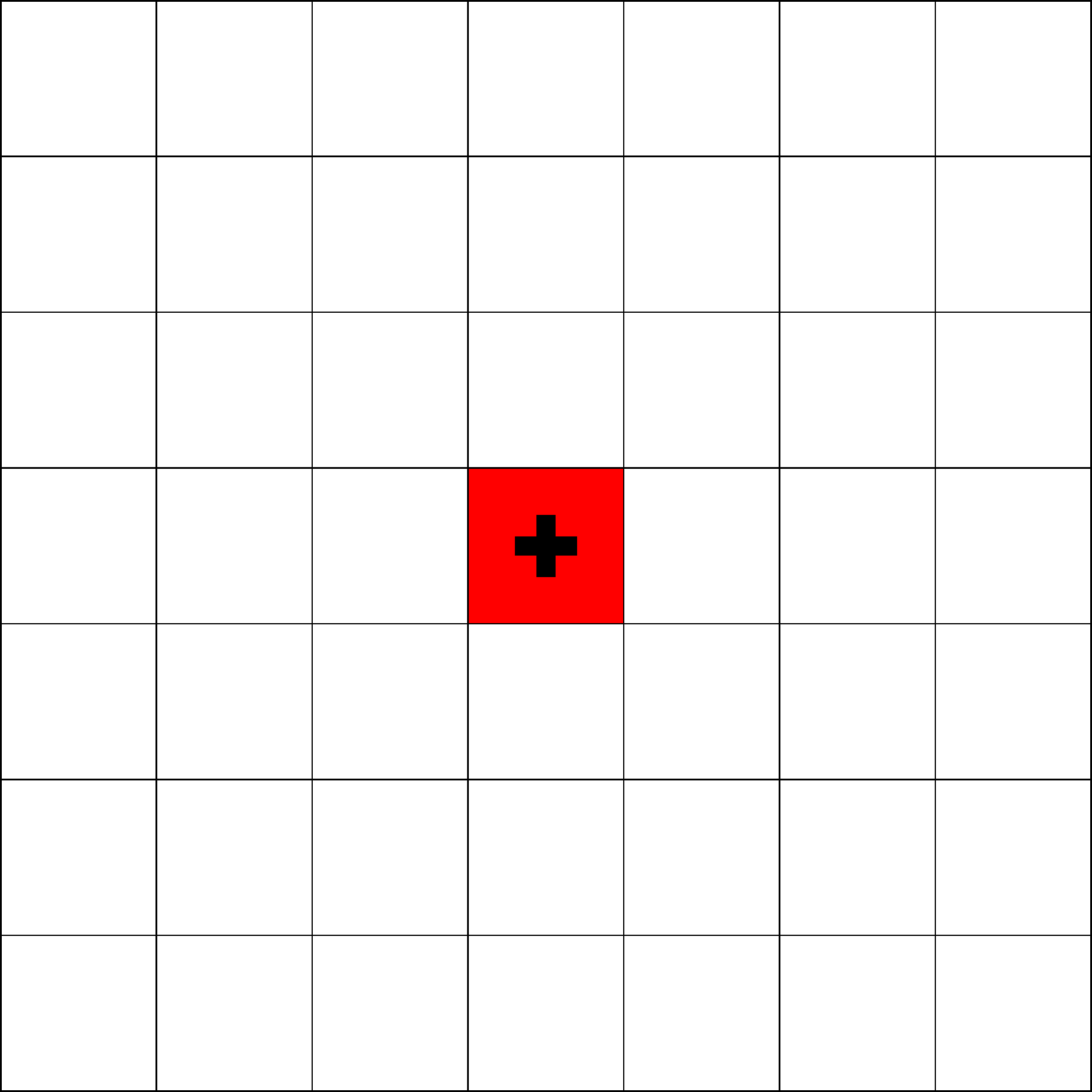}
\caption{$\tiling_{red}$}
\end{subfigure}
\caption{The central fragments of $\tiling_{yellow}$, $\tiling_{med}$ and $\tiling_{red}$.}
\label{fig:broken_triangular_inequality_a}
\end{figure}

The tilings $\tiling_{yellow}$ is defined as follows: there is a $yellow$ tile on the origin (the origin is at the centre of the tile), and all the other tiles are $white$.\\
The tiling $\tiling_{med}$ is defined as follows: there is a $white$ tile on the origin, a $yellow$ tile north of the origin, a $red$ tile east of the origin and all the other tiles are are $white$.\\
The tiling $\tiling_{red}$ is defined as follows: there is a $red$ tile on the origin, and all other tiles are $white$.\\
See Figure \ref{fig:broken_triangular_inequality_a}.

We have $\inf\mathrm{r}_1(\tiling_{yellow}, \tiling_{med}) = 2$ by taking as $P_0\in\tiling_{red}[[\ball{1/2}]]$ and $P_1 \in \tiling[[\ball{1/2}]]$ the patch consisting of the $yellow$ tile and a $white$ tile below, and $\vec{x}=(0,-1)$.\\
Similarly $\inf\mathrm{r}_1(\tiling_{med},\tiling_{red}) = 2$ by taking as $P_0$ and $P_1$ the patch consisting of a $white$ tile and the $red$ tile.\\
However $\inf\mathrm{r}_1(\tiling_{yellow}, \tiling_{red}) = +\infty$ because there is no $r>0$ such that there exist $P_0\in \tiling_{yellow}[[\ball{1/r}]]$, $P_1 \in \tiling_{red}[[\ball{1/r}]]$, $x\in\ball{r}$ and $P_0 = P_1 + \vec{x}$, indeed any such $P_0$ must contain the $yellow$ tile which means that $P_1$ must contain a $yellow$ tile but $\tiling_1$ contains no $yellow$ tile so it is impossible.
\end{example}

The main problem is that when $r>1$ the radius $1/r$ is smaller than the translation vector, which allow for pathological behaviour.

\begin{definition}[Strong distances on geometrical tilings]
\label{def:distance_strong}
Given two tilings $\tiling_0,\tiling_1 \in \fullshift$, we define the distances $\dsa$ and $\dsb$ between $\tiling_0$ and $\tiling_1$ as the minimum between $1$ and the infimum of the radiuses of translation equivalence \ie

\begin{align*}
\dsa(\tiling_0,\tiling_1)&:= \min\left(1, \inf \radiuses{1}{\tiling_0,\tiling_1}\right)
= \inf\left( \left\{ 1 \right \} \cup \radiuses{1}{\tiling_0,\tiling_1}\right)\\
&= \inf \{1\} \cup \{r>0|\, \exists \vec{x}\in \ball{r}, P_0 \in \tiling_0[[\ball{1/r}]], P_1 \in \tiling_1[[\ball{1/r}]],\\
&\hspace{8cm}P_0 = P_1 +\vec{x}\}\\
\dsb(\tiling_0,\tiling_1)&:= \min\left(1, \inf \radiuses{2}{\tiling_0,\tiling_1}\right)
= \inf\left( \left\{ 1 \right \} \cup \radiuses{2}{\tiling_0,\tiling_1}\right)\\
&=\inf \{1\} \cup \{r>0|\, \exists \vec{x}_0,\vec{x}_1\in \ball{r},
(\tiling_0+\vec{x}_0)[\ball{1/r}] = (\tiling_1+\vec{x}_1)[\ball{1/r}]\}
\end{align*}
\end{definition}

\begin{remark}[$1$ or $1/\sqrt{2}$]
The upper bound for the tiling distances can be found with value $2^{-1/2}=1/\sqrt{2}=\sqrt{2}/2$ \cite{robinson2004}, or $1$ \cite{solomyak1997dynamics}. As will be seen later, the definition works for any bound $b$ such that $0< b \leq 1$. Indeed the important thing is that for any $x$ and $y$ such that $0<x,y,(x+y)<b$ we have $1/(x+y) \leq 1/x - y$ which translates to $1/x \geq x+y$. Recall that this bound is necessary, see Examples \ref{example:broken_triangular_inequality_a} and \ref{example:broken_triangular_inequality_b}.
\end{remark}

Let us recall the definition of a distance.
\begin{definition}[Distance]
Let $X$ be a set.
A distance on $X$ is a function $d:X\times X \to \mathbb{R}$ such that :
\begin{enumerate}
\item $d$ is non-negative real valued \ie
\[ \forall x,y \in X,\, d(x,y) \in \mathbb{R} \text{ and } d(x,y) \geq 0 \]
\item $d$ satisfies the identity of indiscernibles \ie
\[ \forall x,y \in X,\, d(x,y) = 0 \Leftrightarrow x = y\]
\item $d$ is symmetric \ie
\[ \forall x,y \in X,\, d(x,y) = d(y,x)\]
\item $d$ satisfies the triangular inequality
\[ \forall x,y,z \in X,\, d(x,z) \leq d(x,y) + d(y,z)\]
\end{enumerate}
\end{definition}


\begin{proposition}
$\dsa$ is a distance on $\fullshift$.
\end{proposition}
\begin{proof}
For this first distance let us prove all four points of the definition of a distance, as will be seen below the only challenging part is the triangular inequality and for the other distances we will only prove the triangular inequality.
\begin{enumerate}
\item $\dsa$ is non-negative and real valued. This is easy to see because for any $\tiling_0$, $\tiling_1$, $\{1\}\cup \radiuses{1}{\tiling_0,\tiling_1}$ is non-empty and included in $[0,+\infty[$ so its infimum is a non-negative real value.
\item $\dsa$ satisfies the identity of indiscernibles. Let is first remark that for any $\tiling$ we have $\dsa(\tiling, \tiling)=0$ because for any $r>0$ is in $\mathrm{r}_1(\tiling,\tiling)$ take $P_0 = P_1 = \tiling[\ball{1/r}]$. Now if $\tiling_0 \neq \tiling_1$ there exists a tile $t$ which belongs to $\tiling_0$ but not to $\tiling_1$, denote $ \vec{z}\in\mathbb{R}^d$ the position of $t$. There are now two cases, either $\tiling_1$ contains no translate of $t$, in which case we have $d(\tiling_0,\tiling_1)\geq 1/\|\vec{z}\|>0$ or $\tiling_1$ contains translates of $t$, denote $\vec{u}$ the smallest vector such that $t+\vec{u} \in \tiling_1$, we now have $d(\tiling_0,\tiling_1)\geq \min(1/\|\vec{z}\|, \|\vec{u}\|)>0$ indeed for any $r\in\mathrm{r}_1(\tiling_0,\tiling_1)$ we have $P_0 = P_1 + \vec{x}$, either $t\in P_0$ then $t-\vec{x} \in P_1 \subset \tiling_1$ so $\|\vec{x}\| \geq \|\vec{u}\|$ \ie $r\geq \|\vec{u}\|$ or $t\notin P_0$ so $\vec{z} \notin \support{P_0}$ \ie $\vec{z} \notin \ball{1/r}$ \ie $r> 1/\|\vec{z}\|$.

\item $\dsa$ is symmetric. Simply remark that $\mathrm{r}_1(\tiling_0,\tiling_1)$ is symmetric, indeed if $r\in\mathrm{r}_1(\tiling_0,\tiling_1)$ then there exists $P_0 \in \tiling_0[[\ball{1/r}]]$,$P_1\in \tiling_1[[\ball{1/r}]]$, $\vec{x} \in \ball{r}$ such that $P_0 = P_1 + \vec{x}$. Now remark that by taking $P_1' = P_1$, $P_0' = P_0$ and $\vec{x}' = - \vec{x}$ we obtain $r\in\mathrm{r}_1(\tiling_0,\tiling_1)$.

\item $\dsa$ satisfies the triangular inequality. Let us take $\tiling_0,\tiling_1, \tiling_2$ such that $\dsa(\tiling_0,\tiling_1)=a$, $\dsa(\tiling_1,\tiling_2)=b$ et $\dsa(\tiling_0,\tiling_2)=c$. Let us show that $c \leq a+b$.

Let us first remark that if $a=0$, $b=0$ or $a+b \geq 1$ the result is trivial, indeed if $a=0$ then $\tiling_0 = \tiling_1$ so $\dsa(\tiling_0,\tiling_2) = \dsa(\tiling_1,\tiling_2)$, and if $a+b \geq 1$ we have $c = \dsa(\tiling_0,\tiling_2)  \leq 1 \leq a+b$ by definition of $\dsa$.

Let us now assume that $0< a, b , (a+b) < 1$. Remark that by Lemma \ref{lemma:strong_inf_min} the $\inf$ is reached for the distances, so there exists $P_0 \in \tiling_0[[\ball{1/a}]]$, $P_1 \in \tiling_1[[\ball{1/a}]]$, $\vec{x}\in \ball{a}$, $P_1'\in\tiling_1[[\ball{1/b}]]$, $P_2' \in \tiling_2[[\ball{1/b}]]$, $\vec{x}' \in \ball{b}$ such that $P_0 - \vec{x} = P_1$ and $P_1' = P_2' + \vec{x}'$. \\
Take $P_1'' := P_1 \cap P_1'$, $P_0'' := P_1'' + \vec{x}$ and $P_2'' := P_1'' -\vec{x}$.

We have $P_0'' = P_2'' + \vec{x} + \vec{x}'$. First remark that $\vec{x}+\vec{x}' \in \ball{a+b}$, so now to prove that $\dsa(\tiling_0,\tiling_2) \leq a+b$ we have to show that $P_0'' \in \tiling_0[[\ball{1/(a+b)}]]$ and $P_1'' \in \tiling_1[[\ball{1/(a+b)}]]$. Since here there are no conditions on $a$ and $b$, they play the same role and we will prove it only for $P_0''$.

What we need to prove actually is $\ball{1/(a+b)} \subseteq \support{P_0''}$. Let us recall that $P_0''$ is defined as $P_0'' := P_1'' + \vec{x} = (P_1 \cap P_1') + \vec{x} = (P_1 + \vec{x}) \cap (P_1' + \vec{x})$ let us now prove that $\ball{1/(a+b)}\subseteq \support{P_1+\vec{x}} \cap \support{P_1'+\vec{x} }$. Remark that $P_1+\vec{x}=P_0$ so $\ball{1/(a+b)} \subset \ball{1/a} \subseteq \support{P_1+\vec{x}}$. For the second part, remark that $\ball{1/b} \subseteq \support{P_1}$ so $\ball{1/b - a} \subset \ball{1/b}+\vec{x} \subseteq \support{P_1'+\vec{x}}$. Now recall that $0<a,b,(a+b)<1$ so we have $1/b \geq a+b $ which implies as $1/(a+b)\leq 1/b -a$ \ie $\ball{1/(a+b)} \subseteq \ball{1/b-a} \subseteq \support{P_1+\vec{x}}$. So $\ball{1/(a+b)} \subseteq \support{P_0''}$. Similarly $\ball{1/(a+b)}\subseteq \support{P_2''}$. Which means that $c \leq a+b$.

\end{enumerate}
\end{proof}

\begin{proposition}
$\dsb$ is a distance of $\fullshift$.
\end{proposition}

\begin{proof}
Here we will only prove the triangular inequality, the real-positivity, identity of indiscernibles and symmetry are very similar to the case of $\dsa$.

$\dsb$ \textbf{satisfies the triangular inequality} \ie
\[\forall \tiling_0, \tiling_1, \tiling_2 \in \fullshift,\ \dsb(\tiling_0,\tiling_2)\leq \dsb(\tiling_0,\tiling_1) + \dsb(\tiling_1,\tiling_2).\]
Let us take $\tiling_0, \tiling_1, \tiling_2 \in \fullshift$. Let us denote $a:= \dsb(\tiling_0,\tiling_1)$, $b:=\dsb(\tiling_1,\tiling_2$ and $c:= a+b$.
Remark that if $c\geq 1$ then the result is trivial because $1$ is an upper-bound on the distance, remark also that if $a=0$ or $b=0$ the result is trivial because of the fact that then $\tiling_0=\tiling_1$ or $\tiling_1 = \tiling_2$. We now assume that $0< a, b, c <1$.\\

Let us now take  $\vec{x}_0,\vec{x}_1 \in \ball{a}$, and $\vec{x}_1',\vec{x}_2' \in \ball{b}$ such that
\begin{align} (\tiling_0 + \vec{x}_0)\cap \ball{1/a} = (\tiling_1 +\vec{x}_1)\cap \ball{1/a}\\ (\tiling_1 + \vec{x}_1')\cap \ball{1/b} = (\tiling_2 + \vec{x}_2')\cap \ball{1/b}
\end{align}
remark that those exist because when the distance is non-zero and less than $1$ the $\inf$ is attained in the distance (see Lemma \ref{lemma:strong_inf_min}).\\
Now add $\vec{x}_1'$ to the first equation, and $\vec{x}_1$ to the second equation. Using Lemma \ref{lemma:trivial_Kpatches} we obtain
\begin{align}
(\tiling_0 + \vec{x}_0 +\vec{x}_1')[ \ball{1/a} + \vec{x}_1'] = (\tiling_1 +\vec{x}_1 + \vec{x}_1')[\ball{1/a} - \vec{x}_1'] \\
(\tiling_1 + \vec{x}_1' + \vec{x}_1)[\ball{1/b} - \vec{x}_1] = (\tiling_2 + \vec{x}_2' +\vec{x}_1)[\ball{1/b} - \vec{x}_1]
\end{align}
Let us  remark that $0< \tfrac{1}{a+b} \leq \tfrac{1}{a} - b$, a simple computation shows that it is equivalent to $\tfrac{1}{a} \geq a + b$ which itself is a direct consequence of $0< a, b , (a+b) < 1$ because then $\tfrac{1}{a} > 1 > (a+b)$. And we also have $0 < \tfrac{1}{a+b} < \tfrac{1}{b} - a$. This means that $\ball{1/c} \subseteq \ball{(1/a) - b}$ and $\ball{1/c}\subseteq \ball{(1/b) - a}$. So using Lemma \ref{lemma:trivial_Kpatches} we obtain
\begin{align}
(\tiling_0 + \vec{x}_0 +\vec{x}_1')[\ball{1/c}] = (\tiling_1 +\vec{x}_1 + \vec{x}_1')[\ball{1/c}]
 = (\tiling_2 + \vec{x}_2' +\vec{x}_1)[\ball{1/c}]
 \end{align}
Now remark that $(\vec{x}_0 + \vec{x}_1'), (\vec{x}_1 + \vec{x}_2') \in \ball{c}$ so overall $\dsb(\tiling_0,\tiling_2)\leq c$ as expected.
\end{proof}

\begin{lemma}[Inf/Min]
\label{lemma:strong_inf_min}
Let $\tiling_0,\tiling_1\in\fullshift$ such that $\dsa(\tiling_0,\tiling_1)=a$ or $\dsb(\tiling_0,\tiling_1)$ with $0<a<1$ then the $\inf$ is attained in the definition of the distance.
\end{lemma}
\begin{proof}
Let $\tiling_0,\tiling_1\in\fullshift$ such that $\dsb(\tiling_0,\tiling_1)=a$ with $0<a<1$.\\
It means that $\inf \radiuses{2}{\tiling_0,\tiling_1} = a$ \ie
\[\forall \epsilon > 0, \exists \vec{x}_0,\vec{x}_1 \in \ball{a + \epsilon}, (\tiling_0 +\vec{x}_0)[\ball{1/(a+\epsilon)}] = (\tiling_1 +\vec{x}_1)[\ball{1/(a+\epsilon)}]\]

In particular there exists two sequences $(\vec{x}_{0,n})_\nN$ and $(\vec{x}_{1,n})_\nN$ such that both
\begin{align*}& \forall \nN,\, \vec{x}_{0,n},\vec{x}_{1,n} \in \ball{a+1/n}\\
&\forall \nN,\, (\tiling_0 + \vec{x}_{0,n})[\ball{1/(a+1/n)}] = (\tiling_1 + \vec{x}_{1,n})[\ball{1/(a+1/n)}]
\end{align*}

By compactness of $\ball{a+1}\times \ball{a+1}$, the sequence $(\vec{x}_{0,n}, \vec{x}_{1,n})_\nN$ has a converging subsequence $(\vec{x}_{0,\alpha(n)},\vec{x}_{1,\alpha(n)}) \to (\vec{x}_0, \vec{x}_1)$. And since $\forall \nN, (\vec{x}_0,\vec{x}_1) \in \ball{a+1/n}$ we have $(\vec{x}_0,\vec{x}_1) \in \bigcap\limits_\nN \ball{a+1/n} = \ball{a}$.

Now let us prove that $(\tiling_0 +\vec{x}_0)[\ball{1/(a)}] = (\tiling_1 +\vec{x}_1)[\ball{1/(a)}]$.
Denote $P_0 := (\tiling_0 +\vec{x}_0)[\ball{1/(a)}]$ and $P_1 := (\tiling_1 +\vec{x}_1)[\ball{1/(a)}]$.
For contradiction let us assume $P_0\neq P_1$. Then there is a tile $t\in P_0\setminus P_1$.\\
Case 1: the interior of the tile $\mathring{t}$ intersects $\ball{1/a}$. In that case there exists $\delta >0$ such that for any $0<\epsilon<\delta$, $\mathring{t}$ intersects the ball $\ball{1/(a+\epsilon)}$. Which is a contradiction because it means that for some $N$, for all $n>N$, $t\in(\tiling_1 + \vec{x}_{1,\alpha(n)})[\ball{1/(a+1/n)}]$, \ie $t\in P_1$.\\
Case 2: the interior of the tile $\mathring{t}$ does not intersect $\ball{1/a}$. This case is actually possible because $P_0$ has to be a patch of tiles (\ie simply connected),
so if taking only the tiles that intersect $\ball{1/a}$ yields a set of tiles with a hole outside of $\ball{1/a}$ then additional tiles are needed to complete it into a patch. In that case there exists a (finite) subset of tiles $X\subseteq P_0$ that intersect $\ball{1/a}$ and that force $t$ \ie $\forall t'\in X,\, \mathring{t'}\cap \ball{1/a} \neq \emptyset$ and for any patch $P$  of $\tiling_0$ if $X\subseteq P$ then $t\in P$. We can then apply the same argument as in case 1 for all the tiles in $X$ and we obtain a contradiction (\ie we obtain that $t\in P_1$).
\end{proof}


\begin{remark}[Difference with tilings of $\mathbb{Z}^d$]
A tiling of $\mathbb{Z}^d$ from alphabet $\mathcal{A}$ is a configuration $c\in A^{\mathbb{Z}^d}$.
In these configurations the position of the tiles are fixed on the grid, only the label can change.
In this setting it is easier to define a distance as
\[ d(c,c'):= 2^{-\inf\{\|x\| , x\in \mathbb{Z}^d, c_x \neq c'_x }, \]

\ie $d(c,c')$ is the inverse of 2 to the power of the coordinate of the first difference between configurations $c$ and $c'$.
This defines a well-known metric which is complete and compact.
\end{remark}

\section{Weak distance on geometrical tilings}
\label{sec:weak}
The idea of the weak distance on tilings is that two tilings are close if on a large ball they are very close in terms of Hausdorff distance \ie the distance between the boundaries of the tilings is small. Weak distances are for example used in \cite{robinson1996, radin1992}.

Let us consider a fixed (finite) tileset $\tileset$ and the full-shift $\fullshift$ \ie the set of tilings with tiles in $\tileset$.

Let $K$ be a compact of $\mathbb{R}^d$, the boundary of $K$ denoted by $\boundary(K)$ is the closure of $K$ (which in the case of a compact, is $K$ itself) minus its interior \ie
\[ \boundary(K) := \overline{K}\setminus \mathring{K}\]
where $\overline{K}$ denotes the closure, and $\mathring{K}$ denotes the interior.

Recall that we denote $\ball{r}$ the closed ball of centre $0$ and radius $r$ in $\mathbb{R}^d$, and $\ball{r}(x)$ the closed ball of centre $x$ and radius $r$ in $\mathbb{R}^d$. We denote $\sphere{r}$ for the sphere of centre $0$ and radius $r$ \ie $\sphere{r} := \boundary \ball{r}$.

\begin{definition}[Hausdorff distance on compacts]
Given two compacts $K_1$ and $K_2$, we can define the Hausdorff distance $H(K_1,K_2)$ in two ways :
\begin{itemize}
\item the first definition is based on the distance between the points of $K_1$ and $K_2$
\begin{align*}
H(K_1,K_2)&:= \max\left(\sup\limits_{x\in K_1}d(x,K_2), \sup\limits_{y\in K_2} d(y,K_1)\right)\\
&:= \max\left(\sup\limits_{x\in K_1}\inf\limits_{y\in K_2} ||x-y||, \sup\limits_{y\in K_2}\inf\limits_{x \in K_1} ||x-y|| \right)
\end{align*}

 \item the second is based on inclusion in a $\epsilon$-neighbourhood
 \[ H(K_1,K_2) := \max\left( \inf\{ \epsilon > 0 | K_1 \subseteq N_\epsilon(K_2)\}, \inf\{ \epsilon > 0 | K_2 \subseteq N_\epsilon(K_1)\}\right)\]
 where $N_\epsilon(K):= \{ x \in \mathbb{R}^d, \exists y \in K, \| x-y\|\leq\epsilon\} = \bigcup\limits_{y\in K}\ball{\epsilon}(y)$.
\end{itemize}
\end{definition}
For more details on the Hausdorff distance, see for example \cite{barnsley1988}[\S 2].

Recall that a patch of tiles is a simply connected finite set of non-overlaping tiles, in particular it is a finite set of compacts.
\begin{definition}[Hausdorff distance on patches of tiles]
Let $P_0$ and $P_1$ be two patches of $\fullshift$. We define the Hausdorff distance between $P_0$ and $P_1$ as
\[ H(P_0,P_1):= \max \left( \max\limits_{t_0 \in P_0} \min\limits_{t_1\in P_1} H(t_0,t_1) , \max\limits_{t_1 \in P_1} \min\limits_{t_0\in P_0} H(t_0,t_1) \right) \]
\end{definition}
Remark that this definition is resembles a lot the first version of the definition of the Hausdorff mesure on compact sets, but here $P_0$ and $P_1$ are finite sets of tiles. This defines a metric of the patches of $\fullshift$ for the same reason as the Hausdorff distance defines a metric on the compact sets of $\mathbb{R}^d$.

\begin{definition}[Boundary and $r$-boundary of a tiling]
Given a tiling $\tiling\in\fullshift$. We define its boundary $\boundary \tiling$ as the union of the boundary of the tiles \ie
\[ \boundary \tiling := \bigcup\limits_{t\in\tiling} \boundary t,\] recall that the tiles are the closure of their interior.
We define its $r$-boundary $\boundary_r(\tiling)$ as the sphere of radius $r$ together with the intersection of the ball of radius $r$ and the boundary of the tiling a \ie
\[ \boundary_r(\tiling) := \sphere{r} \cup \left( \ball{r} \cap \boundary \tiling \right) = \sphere{r} \cup \bigcup\limits_{t\in\tiling} \left(\boundary(t)\cap \ball{r} \right).\]

\end{definition}
With this we can now define two variations of the weak distance on tilings.
\begin{definition}[Weak distance on tilings]
Given two tilings $\tiling_0,\tiling_1 \in \fullshift$ we define the weak tiling distances $\dwc$ and $\dwd$ as follows:

\begin{align*}
 \dwc(\tiling_0,\tiling_1)&:= \inf \{1\}\cup \{ r>0 |\, \exists P_0 \in \tiling_0[[\ball{1/r}]], P_1 \in \tiling_1[[\ball{1/r}]],\, H(P_0,P_1)\leq r\}\\
 \dwd(\tiling_0,\tiling_1)&:= \inf\{1\}\cup\{ r>0 |\ H(\boundary_{1/r}\tiling_0,\boundary_{1/r}\tiling_1)\leq r\}
 \end{align*}
\end{definition}

\begin{remark}[The upper bound]
Here the upper bound given by the $\{1\}$ is not necessary.
Indeed $\dwc$ is naturally bounded by the maximum radius of the tiles in the tileset and $\dwd$ is naturally bounded by $1$.
\end{remark}

\begin{proposition}
$\dwc$ is a distance on $\fullshift$.
\end{proposition}
\begin{proof}
The real positivity and symmetry are clear. We will prove the identity of indiscernibles and the triangular inequality.\\
Identity of indiscernibles: clearly we have $\dwc(\tiling,\tiling)=0$, let us show that if $\tiling_0\neq \tiling_1$ then $\dwc(\tiling_0,\tiling_1)>0$. If $\tiling_0\neq \tiling_1$ there exists a tile $t\in \tiling_0$ such that $t\neq\tiling_1$, in particular we have $\min\limits_{t_1\in \tiling_1} H(t,t_1) = d_t > 0$. Denote $\vec{z}$ the position of $t$. We have $\dwc(\tiling_0,\tiling_1)\geq \min(d_t, 1/\|z\|)$ indeed for any $r,P_0,P_1$ either $t\in P_0$ in which case $H(P_0,P_1) \geq d_t$ or $t\notin P_1$ in which case $\vec{z} \notin \support{P_0}$ so $r>1/\|z\|$.\\
Triangular inequality: take $\tiling_0$, $\tiling_1$, $\tiling_2$ such that $\dwc(\tiling_0,\tiling_1)=a$, $\dwc(\tiling_1,\tiling_2)=b$ and $0<a,b,(a+b)<1$.

We will show that for all $\epsilon>0$ such that $a+b+2\epsilon < 1$, we have $\dwc(\tiling_0,\tiling_2) \leq a+b+2\epsilon$ which by definition of $\dwc$ implies $\dwc(\tiling_0,\tiling_2)\leq a+b$. Let us take $\epsilon>0$, since $\dwc(\tiling_0,\tiling_1)=a < a+\epsilon$, write $a':= a+\epsilon$ there exists $P_0 \in \tiling_0[[\ball{1/a'}]]$, $P_1\in \tiling_1[[\ball{1/a'}]]$ such that $H(P_0,P_1)\leq a'$. Write $b':= b+\epsilon$ and similarly we have $P_1' \in \tiling_1[[\ball{1/b'}]]$ and $P_2'\in \tiling_2[[\ball{1/b'}]]$ such that $H(P_1',P_2')\leq b'$.
Take $P_1'' = P_1\cap P_1'$. Since $P_1'' \subseteq P_1$ and by definition of the Hausdorff distances on patches of tiles, there exists $P_0''\subseteq P_0$ such that $H(P_0'',P_1'') \leq H(P_0,P_1) \leq a'$. Similarly there exists $P_2''$ such that $H(P_1'',P_2'') \leq b'$. So now $H(P_0'',P_2'')\leq a'+b'$.

Let us now prove that $P_0'' \in \tiling_0[[\ball{1/(a'+b')}]]$. We actually show that we can chose $P_0''$ such that $\ball{1/a'} \cap \ball{1/b'-a'} \subset \support{P_0''}$. Indeed $P_1'' = P_1 \cap P_1'$. Assume $P_1 \subseteq P_1'$, then $P_1'' = P_1$ and we have $P_0'' = P_0$ and $\ball{1/a'} \subseteq \support{P_0''}$, now on the contrary if $P_1' \subseteq P_1$ we have $P_1'' = P_1'$ and $\ball{1/b'} \subseteq{P_1'}$ which implies by $H(P_0',P_1')\leq a'$ that $\support{P_0''}\supseteq \ball{1/b'-a'}$ now for the hybrid case it we have the wanted intersection.

Overall $\dwc(\tiling_0,\tiling_1)\leq a'+b'$ since this is for all $\epsilon>0$ we have $\dwc(\tiling_0,\tiling_1)\leq a+b$.
\end{proof}

\begin{proposition}
$\dwd$ is a metric on $\fullshift$.
\end{proposition}
\begin{remark}[boundary and tiles]
We can consider this distance because the tiles are compact which are the closure of their interior, so (given a fixed tileset) the boundary of the tiles determine the tiles.
\end{remark}
\begin{proof}
For the triangular inequality, simply remark that for any tilings $\tiling_0$, $\tiling_1$ and any $R>R'$ we have $H(\boundary_{R'}\tiling_0,\boundary_{R'}\tiling_1) \leq H(\boundary_R\tiling_0, \boundary_R\tiling_1)$. With this small result we can apply a similar strategy as for $\dwc$ to obtain the triangular inequality.

\end{proof}

\begin{remark}[Topology \cite{radin1992}]
In \cite{radin1992} the topology of $\fullshift$ is defined through a countable base $\mathcal{O}$ of open sets on $\fullshift$ defined as :
\[\mathcal{O} := \left( ] P(\prototile_1,\dots \prototile_n, \vec{x}_1, \dots \vec{x}_n)[_{r_1,\dots r_n}\right)_{n\in\mathbb{N},\prototile_i\in \tileset, \vec{x}_i \in \mathbb{Q}^d, r_i \in \mathbb{Q}^{+*} }.\]
Where :
\begin{itemize}
\item $P(\prototile_1,\dots \prototile_n, \vec{x}_1, \dots \vec{x}_n)$ is the non-overlaping set of tiles $\{ \prototile_i + \vec{x}_i, 1\leq i \leq n\}$ (\ie if the $\prototile_i + \vec{x}_i$ overlap, then the $P(\prototile_1,\dots \prototile_n, \vec{x}_1,\dots \vec{x}_n)$ is not defined)
\item $] P(\prototile_1,\dots \prototile_n, \vec{x}_1, \dots \vec{x}_n)[_{r_1,\dots r_n}$ is the set of tilings that contain a patch of $P'=\{ t_i', 1\leq i \leq n\}$ of $n$ tiles such that for each $i$ $H(\prototile_i + \vec{x}_i, t_i')\leq r_i$ where $H$ is the Hausdorff distance.
\end{itemize}
\ie the open sets of $\fullshift$ are defined as the unions of sets of the base $\mathcal{O}$.\\
Note that in \cite{radin1992} they define it with the a countable dense subset $G'$ of the symmetry group of the euclidean space $\mathbb{R}^d$ instead of just the rational translations. For simplicity and because we defined $\fullshift$ as the set of tilings by translates of the prototiles in $\tileset$ we can consider only the translations instead of the whole symmetry group of $\mathbb{R}^d$, and the set of rational translations is countable and dense in the set of translations.\\
Note also that we can consider a single positive rational $r$ instead of a $n$-upple of rational in the definition, it yields the same topology.

This topology is the same as the weak tiling topology defined by $\dwd$, indeed each open set of the base $\mathcal{O}$ contains an open set for $\dwd$, and each open set of $\dwd$ contains an open of the base $\mathcal{O}$.
\end{remark}

\begin{remark}[Coloured tiles]
Contrary to the strong tiling distances, the weak tiling distances are not easily adapted to coloured tiles.
\end{remark}

\begin{lemma}[The strong distance is more discriminant]
The strong tiling distance $\dsb$ is more discriminant than the weak tiling distance $\dwd$. This can be formalized in different ways:
\begin{itemize}
\item  for any two tilings $\tiling_0,\tiling_1 \in \fullshift$ s, we have $\dwd(\tiling_0,\tiling_1) \leq 3 \dsb(\tiling_0,\tiling_1)$
\item any sequence of tilings that converges for $\dsb$ also converges for $\dwd$
\item any open set of for the strong distance $\dwd$ contains an open set of the strong distance $\dsb$. In particular it is true for the open balls $\mathring{\mathbb{B}}$.
\end{itemize}
\end{lemma}
\begin{proof}
Let us prove that if $(\tiling_0+\vec{x}_0)[\ball{1/r}] = (\tiling_1+\vec{x}_1)[\ball{1/r}]$ with $0<r<1$ and with $\vec{x}_0,\vec{x}_1 \in \ball{r}$ then $\dwd(\tiling_0,\tiling_1) \leq 3 r$. Note that for $r>1/3$, $3r>1$ so the result is trivial.
Remark that:
\begin{itemize}
\item $\dwd(\tiling_0, \tiling_0 + \vec{x}_0)=\|\vec{x}_0\| \leq r$ because they differ only by a translation,
\item $\dwd(\tiling_0+\vec{x}_0, \tiling_1+\vec{x}_1) \leq r$ because they coincide on $\ball{1/r}$,
\item $\dwd(\tiling_1+\vec{x}_1, \tiling_1) = \|\vec{x}_1\| \leq r$.
\end{itemize}
We use these inequalities and apply the triangular inequality twice to obtain the expected result.
\end{proof}

%
%
%

\section{A look at the induced topology}
\label{sec:topology}
Let us take a quick look at what open and closed balls of small radius look like.
Let us take a tiling $\tiling \in \fullshift$ and $0< r < 1$.
%

\begin{definition}[The shift action]
$\mathbb{R}^d$ acts on the set of tilings $\fullshift$ by the shift action $\shift$ defined as
$\shift(\vec{x}, \tiling) = \tiling + \vec{x}$.
\end{definition}

\begin{proposition}[Continuity]
The shift action is continuous on $\fullshift$ for both the strong tiling metric and the weak tiling metric.
\end{proposition}

\begin{proof}
Since the strong metric $\dsb$ is more discriminant than the weak metric, we only prove it for the strong metric.\\
Let us recall the definition of a continuous action. An action $f$ of $\mathbb{R}^d$ on $\fullshift$ \ie $f:\mathbb{R}^d\times \fullshift\to\fullshift$ if for any scalar $\vec{x}$ and tilings $\tiling, \tiling'$ such that $f(\vec{x},\tiling)=\tiling'$ and any ball $\mathbb{B}$ of centre $\tiling'$, $f^{-1}(\mathbb{B})$ contains a ball of centre $(\vec{x}, \tiling) \in \mathbb{R}^d\times \fullshift$.

Let $\tiling, \tiling'$ be two tilings, let $\vec{x}$ be a scalar such that $\shift(\vec{x}, \tiling) = \tiling'$.\\ We have $\tiling' = \tiling+\vec{x}$. Now take $0<r<1$, we have :
\[(\vec{x}+\ball{r/2}) \times \mathbb{B}(\tiling, \tfrac{1}{\frac{2}{r} + \|\vec{x}\|}) \subseteq \shift^{-1}(\mathbb{B}(\tiling + \vec{x},r))\]

Let us take $\vec{x}' \in \ball{r/2}$ and $\tiling' \in \mathbb{B}(\tiling, \tfrac{1}{\frac{2}{r} + \|\vec{x}\|})$,\\ let us prove that $\shift(\vec{x} + \vec{x}', \tiling') \in \mathbb{B}(\tiling+\vec{x},r)$.\\
We have :
\begin{itemize}
\item $\shift(\vec{x} +\vec{x}', \tiling') = \tiling' + \vec{x} + \vec{x}'$,
\item $\dsb(\tiling',\tiling)\leq \tfrac{1}{\frac{2}{r}+\|\vec{x}\|}$, \ie there exist $\vec{x}_0,\vec{x}_1 \in \ball{\frac{1}{\frac{2}{r}+\|\vec{x}\|}}$ such that $(\tiling'+\vec{x}_0)[\ball{2/r+\|\vec{x}\|}] = (\tiling+\vec{x}_1)[\ball{2/r+\|\vec{x}\|}]$
\end{itemize}
By adding $\vec{x}+\vec{x}'$ on both sides we have
\[(\tiling'+\vec{x}_0+\vec{x}+\vec{x}' )[\ball{2/r+\|\vec{x}\|}+\vec{x}+\vec{x}'] = (\tiling + \vec{x}_1 + \vec{x} +\vec{x}')[\ball{2/r+\|\vec{x}\|} + \vec{x} + \vec{x}'].\]

We have $\ball{1/r} \subset \ball{2/r +\|\vec{x}\|- r/2 - \|\vec{x}\|} \subset \ball{2/r+\|\vec{x}\|}+\vec{x}+ \vec{x}'$ because $r<1$.
So in particular with $\vec{x}_1':= \vec{x}_1 + \vec{x}' \in\ball{r}$ we have
\[ (\tiling' +\vec{x} + \vec{x'} + \vec{x}_0)[\ball{1/r}] = (\tiling + \vec{x} +\vec{x}_1')[\ball{1/r}]\] so $\tiling'+\vec{x} + \vec{x}' \in \mathbb{B}(\tiling +\vec{x},r)$.
Overall we have the desired inclusion .

\end{proof}
%
%
%
%
%
%
\begin{remark}[Substitutions]
Note that well-defined combinatorial substitutions are also continuous for both the strong and the weak tiling metrics. Note however that it is not trivial for substitutions that are not edge-hierarchic.
\end{remark}

\section{Topological equivalence for FLC subshifts}
\label{sec:topological_equivalence}
\begin{definition}[subshift]
A subshift $X$ is a subset of $\fullshift$ that is closed for the tiling metric and translation invariant \ie $\shift(\mathbb{R}^d,X)=X$.
\end{definition}

\begin{definition}[$K$-subpatches in a subshift]
Given a subshift $X\subseteq \fullshift$ and a compact $K$ of $\mathbb{R}^d$.
We define the $K$-subpatches of $X$, denoted by $X]]K[[$ as the set of patches of the tilings of $X$ that are covered by $K$.

\[X]]K[[:= \{ P |\, \exists \tiling \in X, P\underset{patch}{\subset} \tiling \text{ and } \support{P} \subseteq K \} \]

We define the maximal $K$-subpatches of $X$, denoted by $X]K[$ as the $K$-subpatches of the tilings of $X$ to which a tile cannot be added (and stay a $K$-subpatch) \ie
\[ X]K[:= \{ P \in X]]K[[, \forall \prototile \in \tileset, \forall \vec{x} \in \mathbb{R}^d,  P\cup \{ \prototile + \vec{x} \} \notin X]]K[[ \} \]
\end{definition}

\begin{definition}[FLC] There are many equivalent definitions of FLC, let us state three definitions:
\begin{itemize}
\item a subshift $X$ is said to have finite local complexity (FLC) when for any compact $K$ of $\mathbb{R}^d$, $X]]K[[$ is finite up to translation.
\item a subshift $X$ is FLC when for any $r>0$ there are finitely many patterns of size at most $r$ (\ie covered by a ball of radius $r$) that appear in $X$.
\item  a subshift $X$ is FLC when it contains finitely many 2-tiles patterns.

\end{itemize}
\end{definition}
\begin{definition}[Topological equivalence]
Two distances are topologically equivalent when they define the same topology, in particular they are topologically equivalent when any sequence converges for one distance if and only if it converges for the second distance.
\end{definition}
\begin{proposition}[Topological equivalence]
Let $\tileset$ be a tileset such that the full shift $\fullshift$ has Finite Local Complexity.

All four tiling metrics $\dsa$, $\dsb$, $\dwc$ and $\dwd$ are topologically equivalent, in particular the strong tiling metrics and the weak tiling metrics are topologically equivalent.
\end{proposition}

\begin{proof}
Note that this result is weaker than Theorem \ref{th:metric_equivalence} so we will not give a complete proof here, we will however give the general idea because it will help to understand what happens here.

Let us give the idea of topological equivalence of the strong distance $\dsb$ and the weak distance $\dwd$ on a given FLC full shift $\fullshift$. First remark that $\dwd < \dsb$ so we only need to prove that a sequence that converges for $\dwd$ also converges for $\dsb$. By definition of finite local complexity, for any $r>0$ there are finitely many patterns up to translation in $\fullshift]]\ball{r}[[$. In particular there exists an $\epsilon_r >0$ such that $\epsilon_r<1/r$ and for any $P,P' \in \fullshift]]\ball{r}[[$ if $H(P,P')<\epsilon_r$ then $P$ and $P'$ are identical up to an $\epsilon_r$-translation. Let us now take a sequence $(\tiling_n)_\nN$ that converges to $\tiling$ for $\dwd$. For any $m\in\mathbb{N}$ there exists $N$ such that for all $n>N$, $\dwd(\tiling_n,\tiling) \leq \epsilon_m$, which implies that the central patterns of radius $m$ of $\tiling_n$ and $\tiling$ are the same up to an $\epsilon_m$-translation with $\epsilon_m < 1/m$. Since that is true for all $m$ then the sequence also converges for $\dsb$.
\end{proof}

\begin{figure}[b]
\center
\includegraphics[width=0.5\textwidth]{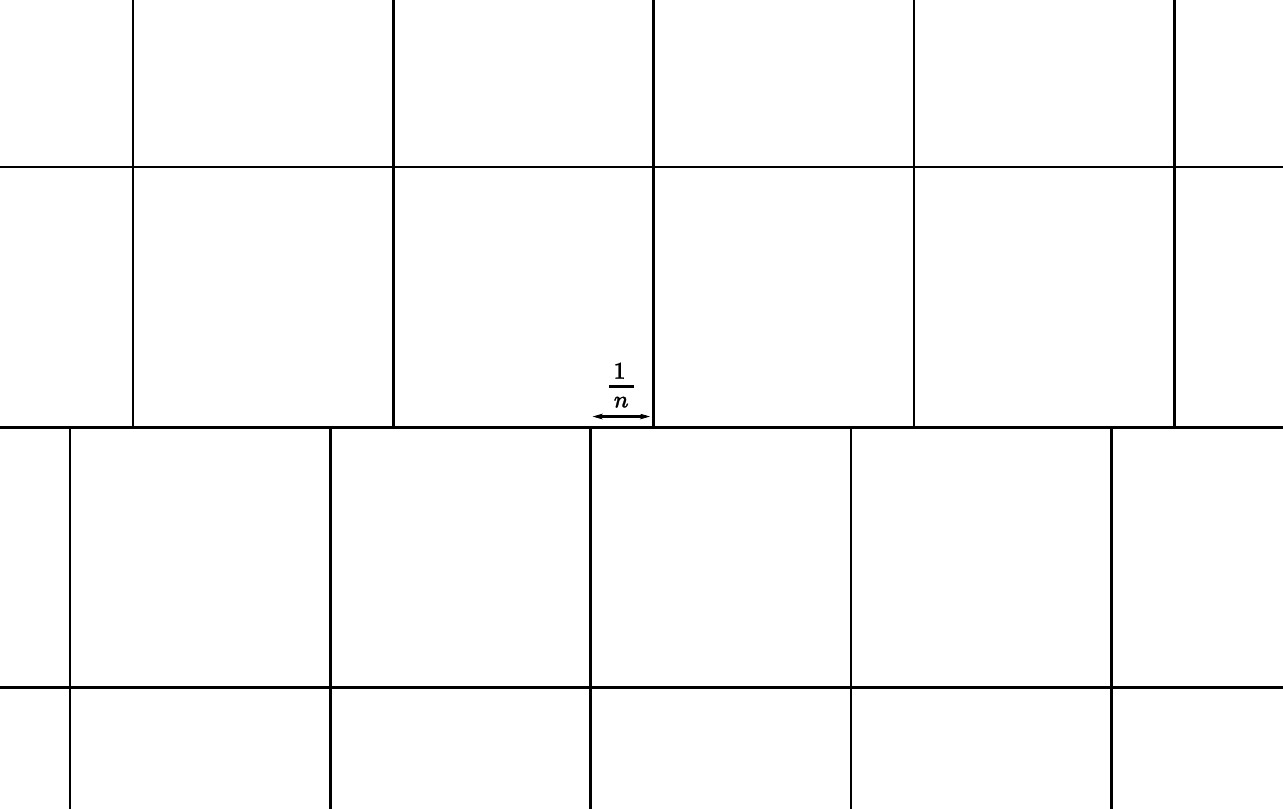}
\caption{The offset tiling $\tiling_n$.}
\label{fig:non-convergence}
\end{figure}

\begin{remark}
Note that if $\fullshift$ does not have finite local complexity, the strong distance and the weak distance are not necessarily topologically equivalent.
See for example the sequence $(\tiling_n)_\nN$ of Example \ref{example:weakconvergence} that converges for $\dwd$ but not for $\dsb$.
\end{remark}
\begin{example}
\label{example:weakconvergence} 
We consider tilings with only one tile: the square unit tile. However we consider tilings that are not necessarily edge-to-edge so that two tiles can share part of an edge.\\
For $n\in\mathbb{N}$ we define the tiling $\tiling_n$ as two half-planes of regular square tilings with an offset $\tfrac{1}{n}$ between the two half-plane (see Figure \ref{fig:non-convergence}).
We also define $\tiling_\infty$ the regular square tiling of the whole plane.

The sequence $(\tiling_n)_\nN$ converges for the weak distances to $\tiling_\infty$ but the sequence does not converge for the strong distances.
\end{example}

\section{Metric equivalence for FLC subshifts}
\label{sec:metric_equivalence}

\begin{definition}[metric equivalence]
Let $X$ be a set and let $d$ and $d'$ be two distances on set $X$. \\
$d$ and $d'$ are called metric equivalent when there exists $\alpha,\beta >0$ such that \[\alpha d \leq d' \leq \beta d.\]
\end{definition}

The tiling metrics $\dsa$, $\dsb$, $\dwc$ and $\dwd$ are metric equivalent on FLC subshifts. Let us start by some simple inequalities.

\begin{lemma}
\label{lemma:simple_inequalities}
Let $\tileset$ be a finite tileset. We have
\begin{enumerate}
\item $\dsb \leq \dsa$
\item $\dwd \leq \dwc$
\item $\dwc \leq \dsa$
\item $\dwd \leq 3 \dsb$
\end{enumerate}
\end{lemma}
\begin{proof}
\begin{enumerate}
\item Let us take $\tiling_0$, $\tiling_1$ $\in \fullshift$ and $0<r<1$ such that $\dsa(\tiling_0,\tiling_1)=r$ since the $\inf$ is reached there exists $P_0 \in \tiling_0[[\ball{1/r}]]$, $P_1 \in \tiling_1[[\ball{1/r}]]$ and $\vec{x} \in \ball{r}$ such that $P_0 = P_1 + \vec{x}$ in particular we have $\tiling_0[\ball{1/r}]\subset P_0$ and $(\tiling_1+\vec{x})[\ball{1/r}] \subset P_1$ so we have $\tiling_0[\ball{1/r}] = (\tiling_1 + \vec{x})[\ball{1/r}]$ \ie $\dsb(\tiling_0,\tiling_1)\leq r$.

\item Let us take $\tiling_0$, $\tiling_1$ $\in \fullshift$ and $0<r<1$ such that $\dwc(\tiling_0,\tiling_1)=r$. Since the $\inf$ is reached, there exists $P_0 \in \tiling_0[[\ball{1/r}]]$ and $P_1 \in \tiling_1[[\ball{1/r}]]$ such that $H(P_0,P_1)\leq r$. Let us now show that $H(\boundary_{1/r}\tiling_0,\boundary_{1/r}\tiling_1)\leq r$.
Recall that $\boundary_{1/r}\tiling_0 = \sphere{1/r}\cup\left(\bigcup\limits_{t\in\tiling_0} \boundary t \cap \ball{1/r}\right)$. Remark also that since there are finitely many tiles that have a non-empty intersection with $\ball{1/r}$, $\boundary_{1/r}\tiling_0$ is a compact as it is a finite union of compacts.
Let us now take $x\in \boundary_{1/r}\tiling_0$, if $x\in\sphere{1/r}$ then $d(x,\boundary_{1/r}\tiling_1)=0$, if $d(x,\sphere{1/r})\leq r$ then $d(x,\boundary_{1/r}\tiling_1)\leq r$ and if $d(x,\sphere{1/r})>r$ then since there exists a tile $t_0\in P_0$ such that $x\in t_0$ and since $H(P_0,P_1)<r$ there exists a tile $t_1\in P_1$ such that $H(t_0,t_1)<r$ also since $x$ is in the boundary of $t_0$ and since $t_0$ and $t_1$ are tiles (\ie compacts which are the closure of their interior) there exists $x'\in \boundary t_1$ such that $d(x_0,x_1)<r$. 
Since $x_1$ is in $\boundary t_1$ and is at distance less than $r$ from $x_0$ (which itself is at distance more than $r$ of $\sphere{1/r}$) then $x_1$ is in $\boundary_{1/r}\tiling_1$ so $d(x_0,\boundary_{1/r}\tiling_1)\leq r$. \\
Now we can do symmetrically and by taking the sup and max we have $H(\boundary_{1/r}\tiling_0,\boundary_{1/r}\tiling_1)\leq r$ which means $\dwd(\tiling_0,\tiling_1)\leq r$.

\item This one is trivial, simply remark that $P_0 = P_1 + \vec{x}$ implies $H(P_0,P_1)\leq \| \vec{x}\|$.

\item This last one can be summed up to $\dwd(\tiling_0,\tiling_1) \leq \dwd(\tiling_0, \tiling_0 + \vec{x}_0) + \dwd(\tiling_0 + \vec{x}_0, \tiling_1 + \vec{x}_1) + \dwd(\tiling_1+\vec{x}_1, \tiling_1) \leq 3r$. 
\end{enumerate}
\end{proof}

Actually the strong distances are metrically equivalent.
\begin{lemma}
Let $\tileset$ be a finite tileset, and $\fullshift$ be the full shift.\\
The strong distances $\dsa$ and $\dsb$ are metrically equivalent.
\end{lemma}
\begin{proof}
As given by Lemma \ref{lemma:simple_inequalities} we have $\dsb < \dsa$.\\
Let us now show that  $\dsa < 2 \dsb$.
Remark that when $\dsb \geq 1/2$ this inequality is trivial, because the upper bound on $\dsa$ is 1.\\
Let us take $\tiling_0$, $\tiling_1$ $\in \fullshift$ and $0<r<1/2$ such that $\dsb(\tiling_0,\tiling_1)<r$. Let us show that $\dsa(\tiling_1,\tiling_0)\leq 2r$.\\
We have the existence of $\vec{x}_0,\vec{x}_1 \in \ball{r}$ such that $(\tiling_0+\vec{x}_0)[\ball{1/r}] = (\tiling_1+\vec{x}_1)[\ball{1/r}]$ which we can rewrite as let us write $(\tiling_0+\vec{x}_0)[\ball{1/r}]$ as $P_0 + \vec{x}_0$ with $P_0 = \tiling_0[\ball{1/r}-\vec{x}_0]$ and same for $P_1+\vec{x}_1$.

We have $P_0 = P_1 + (\vec{x}_1 - \vec{x}_0)$ with $\support{P_0}\supseteq \ball{1/r}-\vec{x}_0 \supseteq \ball{1/r - r} \supseteq \ball{1/2r}$, $\support{P_1}\supseteq \ball{1/r}-\vec{x}_1 \supseteq \ball{1/r - r} \supseteq \ball{1/2r}$ and $(\vec{x}_1 - \vec{x}_0)\in \ball{2r}$. So $\dsa(\tiling_0,\tiling_1)\leq 2r$.
\end{proof}

Similarly the weak distances are also metrically equivalent

\begin{lemma}
Let $\tileset$ be a finite tileset, and $\fullshift$ the associated full shift.\\
The weak distances $\dwc$ and $\dwd$ are metrically equivalent on $\fullshift$.
\end{lemma}

\begin{proof}
We already have $\dwd < \dwc$. Let us now prove the existence of $\beta$ such that $\dwc \leq \beta \dwd$.

This is actually a bit tricky because we need to compare the distances between $1/r$-boundaries and the distance between patches.

\end{proof}

Now that we have proven that the strong distances are metrically equivalent and that the weak distances are also metrically equivalent, let us show that for FLC subshifts all the distances are metrically equivalent.

\begin{proposition}
\label{prop:weak-strong-equivalence}
Let $\tileset$ be a tileset, and $X$ be a subshift of $\fullshift$ with finite local complexity.
The strong tiling distance $\dsa$ and the weak tiling distance $\dwc$ are metrically equivalent.
\end{proposition}

Recall that we have $\dwc \leq \dsa$ so we only need to prove the existence of $\beta$ such that $\dsa \leq \beta \dwc$. In order to prove this, we will need a few intermediate lemmas.
%
%
%

\begin{lemma}
\label{lemma:c1}
Let $\tileset$ be a finite tileset.
There exists $c_1>0$ such that for any two tiles $t_0$, $t_1$ (which up to translation are in \tileset), if $H(t_0,t_1)\leq c_1$ then there exists $\vec{x}\in\mathbb{R}^d$ such that $t_0 = t_1 + \vec{x}$.
\end{lemma}

\begin{proof}
Since the set of tiles is finite up to translation there is a positive minimal Hausdorff distance between any two different tiles up to translation. So with $c_1>0$ the minimal distance between two different tiles the previous statement works. Indeed if $H(t_0,t_1)<c_1$ then $t_0$ and $t_1$ are identical up to translation \ie there exists $\vec{x}\in\mathbb{R}^d$ such that $t_0 = t_1 + \vec{x}$ and moreover we have $|\vec{x} = H(t_0,t_1)$.
\end{proof}

\begin{lemma}
\label{lemma:c2}
Let $\tileset$ be a finite tileset, and let $X$ be a subshift of $\fullshift$ with finite local complexity.\\
There exists $c_2>0$ such that for any two two-tile patches $P_0,P_1 \in \mathcal{L}_2(X)$, if $H(P_0,P_1) \leq c_2$ then there exists $\vec{x}\in\mathbb{R}^d$ such that $P_0 = P_1 + \vec{x}$.
\end{lemma}

\begin{proof} By definition, since $X$ has Finite Local Complexity, there are finitely many two tile patterns up to translation. So the same proof works.
\end{proof}

\begin{lemma}
\label{lemma:c}
Let $\tileset$ be a finite tileset, and let $X$ be a subshift of $\fullshift$ with finite local complexity.\\
There exists $c>0$ such that for any two patches $P_0,P_1 \in \mathcal{L}(X)$, if $H(P_0,P_1)\leq c$ then there exists $\vec{x}\in\mathbb{R}^d$ such that $P_0 = P_1 + \vec{x}$, note that here $\vec{x} = H(P_0,P_1)$.
\end{lemma}

\begin{proof}
Take $c:= \min(c_0,c_1,c_2)$ where $c_0$ is a third of the minimal inner diameter of a tile, $c_1$ is given by Lemma \ref{lemma:c1} and $c_2$ given by Lemma \ref{lemma:c2}.

Now take $P_0, P_1$ such that $H(P_0,P_1)< c$.
First remark that since $H(P_0,P_1)\leq \min(c_0,c_1)$ there is a bijection $f$ from the tiles of $P_0$ to the tiles of $P_1$ such that for any tile $t\in P_0$, there exists $\vec{x}$ such that $f(t) = t+\vec{x}$. The existence of $f$ is given by $H(P_0,P_1)\leq c_1$ and the bijection is due do $H(P_0,P_1) \leq c_0$.

Now remark that since $P_0$ is a patch, if $\vec{x}$ is not uniform then there are two neighbour tiles $t$ and $t'$ in $P_0$ such that $f(t) = t+\vec{x}$ and $f(t') = t' + \vec{x}'$ with $\vec{x}\neq \vec{x}'$, however $H(\{t,t'\}, \{f(t), f(t')\}) \leq c \leq c_2$ so this is a contradiction.

Hence $\vec{x}$ is uniform \ie $P_0 = P_1 + \vec{x}$.

\end{proof}

\begin{proof}[Proof of Proposition \ref{prop:weak-strong-equivalence}.]
Recall that we have $\dwc \leq \dsa$ (Lemma \ref{lemma:simple_inequalities}).
Let us take $c$ of Lemma \ref{lemma:c}. Assume $c\leq 1$, otherwise simply replace $c$ by 1.

Let us first remark that for any two tilings such that $\dwc(\tiling_0,\tiling_1) = d \leq c$ we have $\dsa(\tiling_0,\tiling_1) \leq d$.  Recall that the $\inf$ is attained in the distances, so there exist $P_0 \in \tiling_0[[\ball{1/d}]]$, $P_1 \in \tiling_1[[\ball{1/d}]]$ such that $H(P_0,P_1)\leq d \leq c$, by Lemma \ref{lemma:c} we have $P_0 = P_1 + \vec{x}$ with $|\vec{x}|=H(P_0,P_1) \leq d$ so $\dsa(\tiling_0,\tiling_1) \leq d$.

Now remark that since $\dsa \leq 1$ by definition, we have $\dsa \leq \tfrac{1}{c} \dwc$. Indeed for tilings such that $\dwc(\tiling_0,\tiling_1) \geq c$ the inequality trivially holds because $\dsa \leq 1$, for tilings such that $\dwc(\tiling_0,\tiling_1) \leq c$ we have $\dsa(\tiling_0,\tiling_1) \leq \dwc(\tiling_0,\tiling_1) \leq \tfrac{1}{c} \dwc(\tiling_0,\tiling_1)$.
\end{proof}

\section{Compactness}
\label{sec:compactness}
In this section we present compactness results. Note that the first result is very well known and attributed to \cite[\S 2]{rudolph1989} however in that article the term "compactness" is not used and the proof is kind of obscure so for the sake of completeness we will provide a proof.


\begin{theorem}[FLC subshift are compact]
Let $\tileset$ be a finite set of tiles of $\mathbb{R}^d$. Let $X$ be a subshift of $\fullshift$ with Finite Local Complexity.\\
$X$ is compact for the strong tiling distances (and weak tiling distances).
\end{theorem}

There are many ways to prove this result, we will use K\H{o}nig's lemma here, but note that a proof based on diagonal subsequence extraction works as well. Note that this proof is similar to the classical proof of compactness of the grid tilings $A^{\mathbb{Z}^d}$ using K\H{o}nig's lemma. Remark that the other classical proof of compactness of the grid tilings is based on Tychonoff's theorem and cannot be adapted to geometrical tilings.

\begin{lemma}[K\H{o}nig]
Let $T=(V,E)$ be an finitely branching infinite tree \ie $V$ is infinite and for each vertex $v\in V$ the degree of $v$ is finite.\\
There exists an infinite ray $(v_n)_{n\in\mathbb{N}}$ in the tree, \ie $\forall n, \{v_n,v_{n+1}\} \in E$.
\end{lemma}

\begin{proof}[Proof of FLC subshifts are compact]
This proof is decomposed in two parts : first "fix" an origin point and second use K\H{o}nig's lemma in the same way as on $\mathbb{Z}^d$.

Let $\tileset$ be a finite set of pointed prototiles \ie each prototile $t$ in the tileset has a centre (for example the barycentre of the tile if the tile is convex) which is a point $x_t$ in its interior.
Let $X$ be a subshift of $\fullshift$ with Finite Local Complexity.

To prove that $X$ is compact, let us prove that it is sequentially compact.

Let $(\tiling_n)_\nN$ be a sequence of tilings in $X$. First let us for each tiling $\tiling_n$ we define a central centre $x_n$ as the centre of a tile nearest to the origin of $\mathbb{R}^d$, in case of equidistance chose one arbitrarily.
Let us remark that since the tileset is finite, the tiles have a maximum diameter $D$ so $x_n$ is at distance at most $D$ from the origin.
This implies that the sequence $(x_n)_\nN$ is in $\ball{D}$ so it has a converging subsequence $(x_{\alpha(n)})_\nN$ that converges to some $x_\infty \in \ball{D}$.

We define the centred sequence of tilings $(\tiling_n')_\nN$ as $\tiling_n' := \tiling_{\alpha(n)} - x_{\alpha(n)}$. These are called centred tilings because they all have a centre of tile on the origin of $\mathbb{R}^d$.

Note that the convergence of the centred sequence $(\tiling_n')_\nN$ (resp. of a subsequence of the centred sequence) is equivalent to the convergence of the sequence $(\tiling_\alpha(n))_\nN$ (resp. of a subsequence).

We now define the finitely branching infinite tree $T=(V,E)$ on which to apply K\H{o}nig's lemma. Let $V$ be a root and an infinite union of non-empty sets of vertices $V=\{r\} \cup \bigcup\limits{i\in\mathbb{N}}$ of increasing size defined by
\[ V_i := \{P \in X[[\ball{i}]]\ |\ \exists \nN, P = \tiling_n'[\ball{i}] \},\]
\ie the vertices of $V_i$ are the patches $P$ of radius at least $i$ (they cover the ball $\ball{i})$) which appear infinitely many times in the sequence $(\tiling_n')_\nN$ as the central patch $\tiling_n'[\ball{i}]$.
The set of edges is defined as
\[E := \{ (P, P') \ | \exists i \in \mathbb{N},\, P \in V_i,\, P' \in V_{i+1},\, P \subset P'\} \cup \{ (r,P) |\ P \in V_0\},\]
\ie there is an edge between the patches $P$ and $P'$ when they are of consecutive size and $P'$ contains the patch $P$ at its centre.

Note that we have added a root for $T$ to be a tree, otherwise it would be a forest with each patch of size $0$ (tiles) as root of a tree.

Let us now prove that $T$ is indeed a finitely branching infinite tree. We actually prove that each $V_i$ is non-empty and finite. Take $i\in\mathbb{N}$
\begin{itemize}
\item  $V_i$ is finite. Indeed if $P\in V_i$ then there exists $n$ such that $P = \tiling_n'[\ball{i}]$ which implies that $\ball{i} \subset \support{P} \subset \ball{i+D_\tileset}$ where $D_\tileset$ is the maximum diameter of a tile in the tileset. However by hypothesis $X$ has finite local complexity (FLC) which means that there are finitely many patches $P$ up to translation that are covered by $\ball{i+D_\tileset}$. Now recall that the tilings $\tiling_n'$ are all centred on the origin in the sense that the tiles are pointed and there is a centre of tile at the origin of the space $\mathbb{R}^d$. Remark that patches covered by $\ball{i+D_\tileset}$ have a bounded number of tiles (the bound depends on the $\tileset$ and on the area or volume of $\ball{i+D_\tileset}$) which implies a bounded number of possible centred positions. So there are at most finitely many centred patches covered by $\ball{i+D_\tileset}$. Overall there are at most finitely many patches $P$ in $V_i$.
\item $V_i$ is non-empty. Indeed, for any $n$, $\tiling_n'[\ball{i}]$ is in $V_i$.
\end{itemize}
Let us use this to prove that $T$ is a finitely branching infinite tree:
\begin{itemize}
\item $T$ is has no cycle, indeed if there was a cycle in $T$ then in particular there would be two patches $P_1 \neq P_2$ in some $V_i$ and a patch $P'$ in $V_{i+1}$ such that $(P_1,P')\in E \wedge (P_2,P')\in E$, however this implies $P_1 \subset P'$ and $P_2 \subset P'$ however since both $P_1$ and $P_2$ are defined as a smallest patch that covers $\ball{i}$ then it means $P_1 = P_2$ which is a contradiction. Hence there is no cycle in $T$.

\item $T$ is connected, indeed each patch $P\in V_i$ is connected to the root of the tree because for any $1\leq k \leq i$ its central patch of size $\ball{i-k}$ is in $V_{i-k}$ because it appears in $(\tiling_n')_{nN}$ at least as often as $P$.

\item $T$ is finitely branching, indeed each $V_i$ is finite, the root has $|V_0|$ successors, and a vertex $v\in V_i$ has exactly 1 predecessor and at most $|V_{i+1}|$ successors.

\item $T$ is infinite. Indeed $V$ is infinite because it contains an infinite union of non-empty sets.
\end{itemize}

By K\H{o}nig's lemma, there exists an infinite ray $(P_n)_\nN$ in $T$.
So we have an infinite sequence of increasing patches and we can define the limit tiling $\tiling_\infty' := \bigcup\limits_{\nN} P_n$ which is the limit of a subsequence of $(\tiling_n')\nN$.
Indeed take the function $\beta : \mathbb{N} \to \mathbb{N}$ defined as
\[ \beta(n):= \min \{k \in \mathbb{N} | \exists i \in \mathbb{N}, P_n = \tiling_k'[\ball{i}] \}.\]
By definition of the ray $(P_n)_\nN$, $\beta$ is well defined and strictly increasing and $\tiling_{\beta(n)}' \tendsto \tiling_\infty'$.

Now remark that $\tiling_{\beta\circ\alpha(n)} \tendsto \tiling_\infty' + x_\infty$ so $(\tiling_n)_\nN$ has a converging subsequence.
\end{proof}

\begin{theorem}[subshifts with finitely many tiles up to translation are compact for the weak tiling metric]
Let $\tileset$ be a finite set of tiles of $\mathbb{R}^d$.\\
The fullshift $\fullshift$ is compact for the weak tiling metric.
\end{theorem}

\begin{proof}
For this proof we introduce the concept of \emph{corollas} of a tile and \emph{weak limit patch} of a tiling sequence.

Given a tile $t$ in a patch (or tiling) we say that the $n$th corolla of $t$ is the set of tiles that are at adjacency distance at most $n$ from $t$. For example the $0$ corolla of $t$ is $t$ itself, and the $1$ corolla of $t$ is $t$ together with all the tiles that are adjacent to $t$.
We say that the $n$th corolla of $t$ is \emph{complete} in a patch $P$ with $n\geq 1$ when, for any $k\geq 1$, no tile of the $(n-k)$th corolla is a boundary tile.
Remark that since the tileset is finite up to translation and the tiles have non-empty interior, each tile has finitely many neighbours in a tiling of $X_\tileset$ and each $n$th corolla of a tile contains finitely many tiles.

We say that a patch $P$ is a \emph{weak limit patch} of a sequence of tilings $(\tiling_n)_\nN$ when, for all $n$, there exists a finite set of non-overlaping tiles $P_n\subset \tiling_n$ such that $P_n \tendsto P$ in the Hausdorff distance.
Note that the intermediate sets $P_n$ are finite sets of non-overlaping tiles and not necessarily patches (\ie, not necessarily simply connected).

We consider that the tileset $\tileset$ is pointed \ie each prototile has a "centre" in its interior.

Let $(\tiling_n)_\nN$ be a sequence of tiling in $X$.
We construct a diagonal extraction of of subsequences to fix tiles by order of increasing corolla:

\begin{itemize}
\item fixing the central tile: \\
For each tiling $\tiling_n$ take the centre of tile $x_n$ closest to the origin of $\mathbb{R}^d$, in case of equidistance pick arbitrarily from the finite set of equidistant centres. Remark that since the tileset $\tileset$ is finite up to translation, there exist a prototile $t\in \tileset$ such that there are infinitely many $n$ such that the tile of centre $x_n$ is of type $t$. Define the first extraction $\alpha_0$ that fixes that tile type. Now remark that the sequence $x_{\alpha_{0}(n)}$ is bounded and it lies in $\ball{D_\tileset}$ where $D_\tileset$ is the maximum diameter of a tile in $\tileset$. So it has a converging subsequence to some $x_\infty$, call $\alpha_1$ this new subsequence extractor. So $t+x_\infty$ is a limit tile of the sequence $(\tiling_{\beta_0(n)})_\nN$ with $\beta_0 = \alpha_{0}\circ\alpha_{1}$.
\item fixing the $(k+1)$ corolla:\\
We assume that the subsequence $(\tiling_{\beta_k(n)})_\nN$ that fixes a patch $P_k$ in which the $k$th corolla of the central tile $t+x_\infty$ is complete.
Pick a point of $x_\partial$ the boundary of $P_k$, in each tiling $\tiling_{\beta_k(n)}$ pick $x_n$ a centre of tile (outside $\support{P_k}$)  that contains $x_\partial$. With the same argument as in the first item, there exists a limit $x_{\partial\infty}$ and a tile type $t_\boundary$ such that $P_k \cup (t{\boundary}+x_{\partial\infty})$ is a limit of some $\tiling_{\alpha\circ \beta_k(n)}$.
Note that $(t{\boundary}+x_{\partial\infty})$ is not in $P_k$ because the limit centre $x_{\partial\infty}$ is not in the interior of $\support{P_k}$ whereas the centre of the tiles in $P_k$ are in the interior.
Repeat this operation finitely many times until the $(k+1)$th corolla is complete \ie pick a new point $x_\partial$ in the boundary of $P_k$ not already covered and repeat the extraction.
In the end we obtain a weak limit patch $P_{k+1}$, which is a complete $(k+1)$th corolla, and its extraction $\beta_{k+1}$.
\end{itemize}
Now take the diagonal subsequence $(\tiling_{\gamma(n)})_\nN$ where $\gamma$ is defined as $\gamma(n)= \beta_n(n)$, we have with the weak tiling metric $\dwc$, $\tiling_\gamma(n) \tendsto \bigcup\limits_{k\in\mathbb{N}} P_k$.
Indeed, for all $k$, by construction, $P_k$ is a weak limit patch of the tiling.

This convergence is for the weak tiling metric (and not for the strong tiling metric) because the convergence is separate for each tile.
\end{proof}

\section{Completeness}
\label{sec:completeness}

In this section we prove the fact that the tiling metrics are complete. This is well-known \cite{robinson2004}.

\begin{theorem}[Completeness, folk.]
The tiling metrics $\dsa$, $\dsb$, $\dwc$ and $\dwd$ are complete.
\end{theorem}

\begin{proof}
We only prove the completeness of $\dsa$.

Let $\tileset$ be a tileset of $\mathbb{R}^d$, and $\fullshift$ be the fullshift of tilings with tiles in $\tileset$.

Let us prove that the metric $(\fullshift, \dsa)$ is complete.

Let $(\tiling_n)_\nN$ be a Cauchy sequence of tilings.
Define the sequence $(s_n)_\nN$ as
\[s_n := \dsa(\tiling_n,\tiling_{n+1}) + \frac{1}{2^{n+1}}.\]
We assume that $s_n$ is decreasing, for all $n$ $1/s_{n+1} \geq s_n + 1/s_n$ and $\sum_{\nN}s_n < +\infty$, if it is not the case we may pass to a subsequence that satisfies these conditions.

We now construct a sequence of patches $(P_n)_\nN$ and translations $(x_n)_\nN$ as follows.
For $n=0$, we use $\dsa(\tiling_0,\tiling_1) < s_0$ and the definition of $\dsa$ to define a translation $x_0$ and two patches $P_0\in \tiling_0[[\ball{1/s_n}]]$ and $P_1'\in \tiling_1[[\ball{1/s_n}]]$ such that $P_0 + x_0 = P_1'$. Without loss of generality we assume that $P_0$ and $P_1'$ are such that $P_1' = \tiling_1[\ball{1/s_0} \cup (x_0 + \ball{1/s_0})]$ \ie $P_1'$ is the smallest patch of $\tiling_1$ that contains $\ball{1/s_0} \cup (x_0 + \ball{1/s_0})$ and $P_0$ is the smallest patch of $\tiling_0$ that contains $\ball{1/s_0} \cup (-x_0 + \ball{1/s_0})$.

For $n+1$, given the construction until $n$.
By $\dsa(\tiling_{n+1},\tiling_{n+2}) < s_{n+1}$, we define $P_{n+1}\in \tiling_{n+1}[[\ball{1/s_{n+1}}]]$, $P_{n+2}'\in \tiling_{n+2}[[\ball{1/s_{n+1}}]]$ and $x_{n+1} \in \ball{1/s_{n+1}}$ such that $P_{n+1} + x_{n+1} = P_{n+2}'$ and $P_{n+1} = \tiling_{n+1}[\ball{1/s_{n+1}}\cup (-x_n + \ball{1/s_{n+1}})]$.
Note that $P_{n+1}'$ (defined at the previous step) is included in $P_{n+1}$ because, by our hypothesis that $1/s_{n+1}\geq s_n + 1/s_n$, we have  $\ball{1/s_n} \cup (x_n + \ball{1/s_n}) \subset \ball{1/s_{n+1}}$.

Take $r_n = \sum\limits_{k=n}^{+\infty} x_k$. Since $\sum s_n$ is finite and $\|x_n\|\leq s_n$ this sum converges. The sequence $(r_n + P_n)_{\nN}$ is increasing for the inclusion, indeed $r_n + P_n = r_{n+1} + x_n + P_n = r_{n+1} + P_{n+1}' \subset r_{n+1} + P_{n+1}$.
So the limit tiling $\tiling_\infty$ is the limit of the sequence $(\tiling_n)_\nN$ with \[ \tiling_\infty := \bigcup\limits_\nN r_n + P_n. \]

Remark now that since the sequence $(\tiling_n)_\nN$ is Cauchy, then even if a subsequence was extracted at some step of the construction, then $\tiling_\infty$ is still the limit of the full sequence.
\end{proof}

\bibliographystyle{alpha}
\bibliography{biblio_topology}

\end{document}